\documentclass[aps,twocolumn,superscriptaddress,showkeys]{revtex4-2}

\usepackage{microtype}%Afinar
\usepackage{makecell} % Para forzar saltos de línea en las celdas
\usepackage{graphicx}% Include figure files
\usepackage{dcolumn}% Align table columns on decimal point
\usepackage{bm}% bold math
\usepackage{caption} %Para colocar el caption de las imágenes
\usepackage{subcaption} %Para colocar el caption de las imágenes
\usepackage{mathtools} % Para usar herramientas mathematics
\usepackage{tensor} % Para usar herramientas tensoriales
\usepackage[hidelinks]{hyperref} % Para crear enlaces
\usepackage{amssymb}%Símbolos mathematics
\usepackage{mathtools}%Más simbolos
\usepackage{mathrsfs}%Más símbolos
\usepackage{amsmath} % Para usar la matriz y los símbolos de derivadas parciales
\usepackage{xcolor}%Paquete para el color
\usepackage{array} %Para las tablas
\usepackage{booktabs}%Otra para tablas
\usepackage{enumitem}
\usepackage{soul}%Resaltar
\usepackage[toc,page]{appendix}

\usepackage{multirow}

\begin{document}

%\title{Is the existence of a dressed singularity by throat of a wormhole feasible?}
\title{Axially symmetric wormholes}

\author{I. A. Sarmiento-Alvarado}
\email{Corresponding author: ignacio.sarmiento@cinvestav.mx}
\author{Leonel Bixano}
\email{leonel.delacruz@cinvestav.mx}
\author{Tonatiuh Matos}
\email{tonatiuh.matos@cinvestav.mx}
\affiliation{Departamento de F\'{\i}sica, Centro de Investigaci\'on y de Estudios Avanzados del Intituto Politécnico Nacional, Av. Intituto Politécnico Nacional 2508, San Pedro Zacatenco, M\'exico 07360, CDMX.
}

\date{\today}

\begin{abstract}
    In this work, we derive an exact static, axially symmetric, nonrotating vacuum solution of the Einstein field equations depending on three constant parameters: the throat radius \(r_0\), a parameter \(q\), which is closely associated with the Komar mass, and a parameter \(s\), which determines the axial/polar structure of the geometry. For \(s=0\), the asymptotically flat wormhole sector appears, and the solution reduces to the Gibbons–Volkov ring-wormhole geometry, while the ring conical excess of the extended geometry is removed from the final physical manifold by the cut-and-paste construction. For \(s\neq0\), the solution exhibits a stronger axial singularity instead of a finite axial conical defect.
    
    We employ the cut-and-paste construction to generate wormhole geometries from this solution for $q \neq 0$ and \(s=0\). In addition, we perform a detailed analysis of the embedding diagrams, the wormhole throat, the occurrence and structure of trapped surfaces, the behavior of geodesics, the associated tidal forces, the Petrov algebraic classification, the Newman–Penrose spin coefficients, the energy conditions, and the corresponding invariant conserved charges.
\end{abstract}

\keywords{Wormholes, Surface of revolution , Chiral equation, Surface stress-energy tensor}

\maketitle

%%%%%%%%%%%%%%%%%%%%%%%%%
\section{Introduction}  %
%%%%%%%%%%%%%%%%%%%%%%%%%

Wormholes are among the most emblematic solutions of general relativity: spacetime configurations whose global topology allows two distant regions to be joined by a throat. Although no observations have yet confirmed their existence, they have not been ruled out in a fully model-independent way. The issue is therefore usually framed in terms of specific geometric and energetic criteria that determine when wormholes can arise and which observables might distinguish them from ordinary compact objects. The most emblematic work was done by Morris and Thorne, who emphasized the throat’s geometry, an operational definition of traversability based on tidal-force limits, and the tight link between wormhole feasibility and the energy conditions \cite{Morris:1988cz}.
In \cite{PhysRev.48.73}, Einstein and Rosen introduced the first wormhole.
Later, Bronnikov and Ellis (BE) \cite{Bronnikov:1973fh,10.1063:1.1666161} derived a wormhole solution by examining the EFE coupled to a scalar field.
Currently, wormholes are studied within General Relativity and its extensions.
Wormhole solutions have been obtained in higher dimensions \cite{Clement1984, LU2008511, Dzhunushaliev:1998rz, Deshpande:2022zfm, Sarmiento-Alvarado:2025msh}, in Einstein-Gauss-Bonnet theory \citep{PhysRevD.46.2464,PhysRevD.77.089903, PhysRevD.78.024005}, in Einstein-Maxwell-dilaton theory \cite{ Bixano:2025jwm,Bixano:2025bio}, in $f(R)$ theory \citep{Lobo:2009ip, Pavlovic2015, DeFalco:2021ksd}, in Brans–Dicke gravity \cite{EIROA20092399, PhysRevD.51.2011} and in scalar-tensor gravity \cite{PhysRevD.94.024011, Franciolini2019}.

A key structural obstacle is topological censorship: in asymptotically flat, globally hyperbolic spacetimes satisfying the null energy condition, causal curves from past to future null infinity cannot access nontrivial topology \cite{Friedman:1993ty}. This does not forbid wormholes as solutions, but it strongly restricts classical traversability under standard energy assumptions. As a result, most work proceeds along two lines: introducing matter sources that controllably violate the energy conditions or using geometric constructions that localize the nontrivial topology so that its physical and causal effects can be cleanly analyzed. A standard example of the second scenario is the cut-and-paste construction of Lorentzian wormholes, where two manifolds are glued across a timelike hypersurface and the supporting stress-energy is confined to a thin shell fixed by the Israel junction conditions \cite{Israel:1966rt,Visser:1989kg}. Beyond providing a clean separation between bulk geometry and throat dynamics, this framework also aligns naturally with the wormhole-based formulation of cosmic censorship, which refines the standard censorship principle by insisting that any curvature singularities remain causally unreachable for physically relevant observers and geodesics \cite{axioms14110831,Bixano:2025qxp}.

From this perspective, it becomes simpler to study families of exact wormhole geometries, explicitly guided by the diagnostic criteria most relevant to traversability and censorship. Let us begin with the solution presented in \cite{Gibbons:2016bok}, where the authors, Gibbons and Volkov, propose a method for constructing traversable wormholes.
They obtained the Ring wormhole:
\begin{equation}
\label{ring wormhole}
    ds^2
    = - e^{2 \lambda \Phi} dt^2
    + \frac{
        \Delta^{1 + \lambda^2} [
            dr^2
            + \Delta_r d\theta^2
        ]
        + \Delta_r \sin^2\theta d\phi^2
    }{e^{2 \lambda \Phi}} ,
\end{equation}
where $\Phi = \arctan \tfrac{r}{r_0}$, $\Delta_r = r^2 + r_0^2$, $\Delta_\theta = r^2 + r_0^2 \cos^2 \theta$, $\Delta = \frac{\Delta_\theta}{\Delta_r}$, $r_0$ is a positive constant, $\lambda \in \mathbb{R}$ is a constant parameter, $r \in [ 0, \infty )$ and $\theta \in I_\theta = [ 0, \pi ]$.

The above metric has a ring singularity of radius $r_0$.
When $\lambda = 0$, the wormhole throat is located at $r = 0$ and can be viewed as a disk.
In this case, the singularity encloses the disk.
However, the metric \eqref{ring wormhole} has not been studied for $\lambda \neq 0$.

In \cite{10.1063/1.1705005}, Zipoy obtains the ring wormhole considering a vacuum spacetime.
They start with the completely general static cylindrically symmetric metric, express it in oblate spheroidal coordinates, and then solve the EFE in these coordinates.
The solutions depend on a constant parameter $l$, and the Newtonian potential is given by a linear combination of Legendre polynomials of the first kind.
The case $l = 0$ corresponds to the ring wormhole solution.
Voorhee introduces in \cite{PhysRevD.2.2119} a construction of static, axially symmetric solutions in general relativity based on Weyl coordinates, together with a clear physical interpretation of them.
In this way, these metrics can be characterized in terms of gravitational potentials.
As an example, he assumes that the Newtonian potential is similar to that of a ring wormhole, but he uses the arccotangent function instead of the arctangent, which are related by the identity $\operatorname{arccot}  x = \tfrac{\pi}{2} - \arctan x$ for all $x \in \mathbb{R}$.
Bronnikov and Fabris \cite{Kirill_A_Bronnikov_1997} generalize the ring wormhole to $D$-dimensional Einstein dilaton gravity.
They consider a $D$-dimensional pseudo-Riemannian manifold $M^4 \times V_1 \times \cdots \times V_n$, where $M^4$ is the 4-dimensional spacetime, and $V_i$ are conformally Ricci-flat manifolds of dimensions $N_i$, with $D = 4 + N_1 + \ldots + N_n$.
To solve the field equations, they assume a static axially symmetric metric for $M^4$, allowing the construction of higher-dimensional wormhole solutions and the analysis of their geometric and physical properties.
These authors obtain the ring wormhole solving the EFE for a static axially symmetric metric.

The flat subspaces method proposed in \cite{Sarmiento-Alvarado2023, Sarmiento-Alvarado2025} to solve the vacuum EFE in higher dimensions allows this metric to be generalized.
Four-dimensional solutions can also be constructed from it.
Through this algebraic method, wormhole solutions in higher dimensions are constructed in \cite{Sarmiento-Alvarado:2025msh}.
Furthermore, this method is used to obtain solutions to the EFE with a non-zero cosmological constant in higher dimensions, including four dimensions \cite{Sarmiento-Alvarado:2026acc}.

In this work, we obtain an exact wormhole solution to the vacuum Einstein field equations (EFE) using the flat subspaces method.
This solution contains the ring wormhole as a particular case.
Our main objective is to investigate the properties of the ring wormhole for $\lambda \neq 0$.

The paper is organized as follows. In Section \ref{section: gen sol}, we derive a general solution to the EFE, and its properties are analyzed in Section \ref{section: struct sols}.
In Section \ref{section: Clasificacion de PETROV}, we determine the Petrov classification and examine the corresponding asymptotic behavior.
Subsequently, in Section \ref{section: Komar cargas}, we compute the invariant conserved charges in the sense of Komar and interpret the parameters $q$ and $s$.
In Section \ref{section: trapped surfaces}, we identify the trapped surfaces and the relevant spin coefficients, which are then analyzed in detail.
In Section \ref{section: throat}, we construct wormhole geometries from the obtained solution.
The associated embedding surfaces are studied in Section \ref{section: embedding surfaces}.
In Section \ref{section: geodesics}, we derive the conditions that ensure the existence of solutions to the geodesic equations.
The surface stress-energy tensor is computed in Section \ref{section: energy conditions}, and it is then used to check the energy conditions.
In Section \ref{section: tidal forces} we examine the tidal forces and determine the safest region for traversing the wormhole.
Finally, in Section \ref{section: conclusions}, we present our concluding remarks.

%%%%%%%%%%%%%%%%%%%%%%%%%%%%%
\section{General solution}  %
%%%%%%%%%%%%%%%%%%%%%%%%%%%%%
\label{section: gen sol}

The Einstein field equations (EFE) in vacuum are equivalent to $R_{\mu \nu} = 0$ for all $\mu, \nu \in \{ 0, \dots, 3 \}$.
If we consider a spacetime endowed with a metric $\hat{g}$ that admits two commutative Killing vectors, then there exists a system of coordinates where the metric has the form
\begin{equation}
    \hat{g}
    = \varrho [ ( dx^1 )^2 + ( dx^2 )^2 ]
    + g_{i j} dx^i dx^j ,
\end{equation}
so that the EFE are reduced to
\begin{align}
\label{chiral eq}
&    ( \upsilon g_{, w} g ^{-1} )_{, \bar w} + ( \upsilon g_{, \bar w} g ^{-1} )_{, w} = 0 ,
\\\label{field eq f}
&   ( \ln \varrho \sqrt{\upsilon} )_{, W} = \tfrac{\upsilon}{2} \operatorname{tr} ( g_{, _W} g^{-1} )^2
    \text{ for } W \in \{ w, \bar w \},
\end{align}
where $g = -\upsilon^{-1} g_{i j}$, 
$\upsilon = \sqrt{ -\det g_{i j} }$ for all $i, j \in \{ 3, 4 \}$, $w = x^1 + i x^2$ and $x^1 = \upsilon$.
The metric components $\varrho$ and $g_{i j}$ depend on $x^1$ and $x^2$.
Consequently, the matrix $g$ also depends on $x^1$ and $x^2$.
Since the determinant of $g$ is $\det g = -1$, $g$ is a symmetric matrix in $SL ( 2, \mathbb{R} )$.

Eq. \eqref{chiral eq} is a nonlinear matrix differential equation known as the chiral equation.
It is solved in \cite{Sarmiento-Alvarado2023, Sarmiento-Alvarado2025} for symmetric matrices in $SL ( n, \mathbb{R} )$, with $n \geq 2$.
Some solutions are given by
\begin{equation}
\label{sol chiral eq}
    g ( w, \bar w )
    = \exp( \zeta ( w, \bar w ) A ) g_0 ,
\end{equation}
where $A$ is a constant matrix in $\mathfrak{sl} ( n, \mathbb{R} )$, $g_0$ is a constant matrix in $\mathcal{I} (A)$ and $\zeta$ is a parameter that satisfies the generalized Laplace equation,
\begin{equation}
\label{Laplace eq}
    ( \upsilon \zeta _{, w} )_{, \bar w} + ( \upsilon \zeta _{, \bar w} )_{, w} = 0 .
\end{equation}
Given that Eq. \eqref{sol chiral eq} is taken into account, the field equation \eqref{field eq f} for the function $\varrho$ becomes
\begin{equation}
\label{field eq f A}
    ( \ln \varrho \sqrt{\upsilon} )_{, W} = \tfrac{\upsilon}{2} \operatorname{tr} A^2 \zeta_{, _W}^2 .
\end{equation}

In Boyer-Lindquist coordinates, $x^1 = \sqrt{\Delta_r } \sin \theta$ and $x^2 = r \cos \theta$, Eq. \eqref{Laplace eq} transforms to
\begin{equation}
\label{Laplace eq Boyer-Lindquist coordinates}
    ( \Delta_r \zeta_{, r} ) _{, r} + \tfrac{1}{\sin \theta} ( \zeta_{, \theta} \sin \theta ) _{, \theta} = 0 .
\end{equation}
We solve Eq. \eqref{Laplace eq Boyer-Lindquist coordinates} considering $\zeta ( r, \theta ) = \zeta_r (r) + \zeta_\theta ( \theta )$, obtaining
\begin{equation}
    \zeta ( r, \theta )
    = \zeta_0
    + \mathscr{C} \ln \upsilon 
    + \tfrac{\mathscr{D}}{r_0} \Phi
    + \mathscr{E} \ln \tan \tfrac{\theta}{2} ,
\end{equation}
where $ \zeta_0, \mathscr{C}, \mathscr{D}, \mathscr{E} \in \mathbb{R}$.
Once we know $\zeta$, we can determine $\varrho$.
To do it, we solve Eq. \eqref{field eq f A} in Boyer-Lindquist coordinates,
\begin{equation}\label{diff eq f Boyer-Lindquist coordinates}
\begin{aligned}
    ( \ln \varrho \sqrt{\upsilon} )_{, r}
&   = \tfrac{\operatorname{tr} A^2}{4} \tfrac{
        r \sin ^2 \theta
    }{\Delta} (
        \zeta^2_{, r} 
        + \tfrac{
            2 \cot \theta \zeta_{, r} \zeta_{, \theta}
        }{r}
        - \tfrac{
            \zeta^2_{, \theta}
        }{\Delta_r}
    ) ,
\\  ( \ln \varrho \sqrt{\upsilon} )_{, \theta}
&   = \tfrac{\operatorname{tr} A^2}{-8} \tfrac{
        \sin 2 \theta
    }{ \Delta } (
    \zeta^2_{, r}
   - \tfrac{
            2 r \tan \theta \zeta_{, r} \zeta_{, \theta}
        }{\Delta_r}
    - \tfrac{
            \zeta^2_{, \theta}
        }{\Delta_r}
    ) .
\end{aligned}
\end{equation}
Then, $\varrho$ is given by
\begin{equation}
    \varrho  ( r, \theta )
    = \mathscr{F} \upsilon^\mathfrak{a}
    \tan^\mathfrak{c} \tfrac{\theta}{2}
    \Delta^{\mathfrak{d} }
    \Xi^\mathfrak{f}
    \exp( \mathfrak{b} \Phi + \mathfrak{e} \Theta ) ,
\end{equation}
where $\Xi = \tfrac{ \sin^2 \theta }{\Delta_\theta}$,
$\Theta = \arctan ( \frac{r}{r_0} \sec \theta )$,
$\mathscr{F}$ is a positive constant,
$\mathfrak{a} = \tfrac{ \mathscr{C}^2 }{4} \operatorname{tr} A^2 - \tfrac{1}{2}$,
$\mathfrak{b} = \frac{\mathscr{C} \mathscr{D}}{2 r_0} \operatorname{tr} A^2$,
$\mathfrak{c} = \tfrac{ \mathscr{C} \mathscr{E} }{2} \operatorname{tr} A^2$,
$\mathfrak{d} = \frac{\mathscr{D}^2}{8 r_0^2} \operatorname{tr} A^2$,
$\mathfrak{e} = \frac{\mathscr{D} \mathscr{E}}{2 r_0} \operatorname{tr} A^2$ and
$\mathfrak{f} = \frac{ \mathscr{E}^2 }{8}\operatorname{tr} A^2$.

We construct an exact solution to the EFE by considering the matrix $A = \operatorname{diag} ( 1, -1 ) \in \mathfrak{sl} ( 2, \mathbb{R} )$.
Given that $g_0$ belongs to $\mathcal{I} (A)$, it has the general form $g_0 = \operatorname{diag} ( -c_0, c_0^{-1} )$, where $c_0$ is a positive constant.
Furthermore, we set $\zeta_0 = 0$, $\mathscr{C} = -1$, $\mathscr{D} = q r_0$, $\mathscr{E} = 2 s$, $\mathscr{F} = 1$ and $c_0 = 1$.
Therefore,
\begin{equation}
\label{metric}
    \hat{g}
    = - f dt^2 
    + \frac{
        \Delta^\ell  h (
            dr^2
            + \Delta_r d\theta^2
        )
        + \Delta_r \sin^2 \theta d\phi^2
    }{f} ,
\end{equation}
where $f = \exp( q \Phi ) \tan^{2 s} \tfrac{\theta}{2}$, $h = \Xi^{s^2} \exp( 2 q s \Theta )$, $\ell = 1 + \frac{q^2}{4}$, $q, s \in \mathbb{R}$ are constant parameters.

\section{Structure of the solutions}
\label{section: struct sols}
%In this section, we will study the metric \eqref{metric}.

When $q = s = 0$, the metric \eqref{metric} reduces to
\begin{equation}
    \hat{g}
    = - dt^2
    + \Delta dr^2
    + \Delta_\theta d\theta^2
    + \Delta_r \sin^2 \theta d\phi^2
 ,
\end{equation}
which is the Minkowski metric in oblate spheroidal coordinates.
These coordinates are related to the Cartesian coordinates as $x = \sqrt{ \Delta_r } \sin \theta \cos \phi$,
$y = \sqrt{ \Delta_r } \sin \theta \sin \phi$ and
$z = r \cos \theta$.

The metric \eqref{metric} is also defined for negative values of $r$.
We extend the values of $r$ to the entire $\mathbb{R}$ with the purpose of having the metric describe a wormhole.
The non-negative values of $r$ correspond to Universe 1, while the negative values correspond to Universe 2.

Now, we analyze the asymptotic behavior of the metric \eqref{metric}.
Observe that $h \to 0$ as $r \to \pm\infty$.
This means that $\hat{g}$ is not asymptotically flat when $s \neq 0$.
Now, we assume that $s = 0$.
Hence,
\begin{equation}
\label{asymptotic metric}
    \hat{g} \to
    \frac{
        dr^2
        + r^2 d\Omega^2
    }{e^{ \pm q \frac{\pi}{2} }}
    - e^{ \pm q \frac{\pi}{2} } dt^2
    \text{ as }
    r \to \pm\infty .
\end{equation}
Under scaling transformations: $r \to e^{ \mp q \frac{\pi}{4} } r$ and $t \to e^{ \pm q \frac{\pi}{4} } t$, the metric \eqref{asymptotic metric} changes to
\begin{equation}
    \hat{g} \to
    - dt^2
    + dr^2
    + r^2 ( d\theta^2 + \sin^2 \theta d\phi^2 ) ,
\end{equation}
which is the Minkowski metric in spherical coordinates.

In the remainder of this paper, we consider only the case $s = 0$.
Consequently, $h = 1$ and $f = \exp( q \Phi )$.
Observe that the metric \eqref{metric} reduces to the metric \eqref{ring wormhole} with $q = 2 \lambda$.

The Kreschman invariant, denoted by $K$, of the metric \eqref{metric} is given by
\begin{equation}
    K = - \frac{q^2 r_0^2}{4} \frac{ \Delta_r^{2 \ell - 4} }{ \Delta_\theta^{ 2 \ell + 1 } } ( R + r_0^2 S \cos^2 \theta ) f^2 ,
\end{equation}
where $R (r) = -( 16 + 8 q^2 + q^4 ) r_0^4 + 6 q ( 4 + q^2 ) r_0^3 r - 24 ( 2 + q^2 ) r_0^2 r^2 + 48 q r_0 r^3 - 48 r^4$ and $S (r) = ( 16 - 4 q^2 + q^4 ) r_0^2 + 6 q ( 4 - q^2 ) r_0 r + 12 q^2 r^2$.
The functions $\Delta_r$ and $f$ are positive on $\mathbb{R}$.
However, the function $\Delta_\theta$ is zero at $( r, \theta ) = ( 0, \tfrac{\pi}{2} )$.
Since $l$ is a positive constant, $K$ is singular at $( r, \theta ) = ( 0, \tfrac{\pi}{2} )$.

%%%%%%%%%%%%%%%%%%%%%%%%%%%%%%%%%%%%%%%%%%%%%
\section{Petrov Classification}\label{section: Clasificacion de PETROV}    %
%%%%%%%%%%%%%%%%%%%%%%%%%%%%%%%%%%%%%%%%%%%%%
To identify the Petrov class of the spacetime, we introduced a Newman–Penrose diagonal null tetrad
{\small
\begin{equation*}
\begin{aligned}
\ell
&=
\frac{1}{\sqrt{2}}
\left(
\sqrt{-\,g^{tt}}\;\partial_t
+
\sqrt{g^{rr}}\;\partial_r
\right),\\
n
&=
\frac{1}{\sqrt{2}}
\left(
\sqrt{-\,g^{tt}}\;\partial_t
-
\sqrt{g^{rr}}\;\partial_r
\right),\\
m
&=
\frac{1}{\sqrt{2}}
\left(
\sqrt{g^{\theta\theta}}\;\partial_\theta
+
i\,\sqrt{g^{\phi\phi}}\;\partial_\phi
\right),\\
\bar m
&=
\frac{1}{\sqrt{2}}
\left(
\sqrt{g^{\theta\theta}}\;\partial_\theta
-
i\,\sqrt{g^{\phi\phi}}\;\partial_\phi
\right).
\end{aligned}
\end{equation*}}
Using this tetrad, we applied the usual contractions of the Weyl tensor and subsequently computed the five Newman–Penrose Weyl scalars $\Psi_0,\dots,\Psi_4$:
{\small
\begin{align*}
    &\Psi_0 = - C_{abcd} \, \ell^a m^b \ell
    ^c m^d, \quad
    \Psi_1 = - C_{abcd} \, \ell^a n^b \ell^c m^d, \\
    &\Psi_2 = - C_{abcd} \, \ell^a m^b \bar{m}^c n^d, \quad
    \Psi_3 = - C_{abcd} \, \ell^a n^b \bar{m}^c n^d, \\
    &\Psi_4 = - C_{abcd} \, \bar{m}^a n^b \bar{m}^c n^d,
\end{align*}
}
where $C_{abcd}$ is the Weyl tensor.

In our analysis, all five scalars are listed in Appendix \ref{Newman-Penrose Weyl Escalares}, so we can see that all the scalars $\Psi \neq 0$.
We then formed the usual algebraic invariants
{\small
\begin{equation}
    \mathcal{I} \;=\; \Psi_0 \Psi_4 - 4 \Psi_1 \Psi_3 + 3 \Psi_2^2 \, , 
    \quad
    \mathcal{J} \;=\; \det
    \begin{pmatrix}
    \Psi_4 & \Psi_3 & \Psi_2 \\
    \Psi_3 & \Psi_2 & \Psi_1 \\
    \Psi_2 & \Psi_1 & \Psi_0
    \end{pmatrix}.
\end{equation}
}

These are both nonzero for generic choices of the parameters and coordinates. In addition, we obtain $\mathcal{I}^3 \neq 27 \mathcal{J}^2$, implying that the Weyl tensor is algebraically general (Petrov type I) according to the standard invariant classification \cite{Newman:1961qr,Pirani:1956wr,Cita:ExactSolutions,Zakhary:1997xas}. The explicit formulas for the invariants $\mathcal{I}$ and $\mathcal{J}$ are in the Appendix \ref{Newman-Penrose Weyl Escalares}, instead, in the next subsection we will discuss the asymptotic behaviour of the NPW scalars.

\subsection{Asymptotic Weyl scalars and physical interpretation}

Throughout all subsequent paragraphs, reference will be made to Appendix \ref{Newman-Penrose Weyl Escalares}.

\paragraph{Coulomb term and effective mass.}
The leading decay behavior $\Psi_2\sim r^{-3}$ corresponds to the usual Coulomb-type falloff of the Weyl tensor for asymptotically isolated systems. More precisely, in the standard asymptotic normalization one finds $-\Psi_2\sim -M/r^3+\cdots$ for a pure mass monopole. Comparing this with $\lim_{ r \to \pm\infty } \, \Psi_2$ one can defines an effective monopole mass 
\begin{equation}
  M_{\rm eff}\ \sim\ \pm \,\frac{qr_0}{2}\, f_{\pm \infty},
\end{equation}
it is intimately connected to the Komar mass $M_K$ that will be computed in the next section \ref{section: Komar cargas}. Under a wormhole interpretation, this implies that both asymptotic regions perceive the same mass parameter. At large distances from the throat, the Weyl tensor is thus primarily governed by a Coulomb-type (monopolar) tidal component described by $\Psi_2$, while the higher-order multipole tidal contributions, given by $\Psi_{0,1,3,4}$, diminish more rapidly and can be neglected.

\paragraph{Absence of gravitational radiation.}
A genuinely radiative outgoing part would appear as a leading $\Psi_4\sim r^{-1}$ term at null infinity. However, $\lim_{ r \to \pm\infty } \, \Psi_4$ instead yields $\Psi_4 \sim \mathcal{O}(r^{-5})$, which is far too suppressed to describe a true gravitational-wave flux. This agrees with the fact that the solution is stationary. Similarly, the equality $\Psi_0=\Psi_4$ indicates a symmetry between ingoing and outgoing null directions in the chosen tetrad, rather than signaling actual radiation.

\paragraph{Higher multipoles and axial angular structure.}
The scalars $\Psi_0$ and $\Psi_1$ fall off as $r^{-5}$ and exhibit explicit angular
dependence via $\sin^2\theta$ and $\sin(2\theta)$, which is typical of
axisymmetric higher-multipole (tidal) structure rather than radiative modes.
An important indicator is that the multipolar corrections are reduced by
two additional powers of $r$ compared to the Coulomb term at the infinity:
{\small
\begin{align}
  \frac{|\Psi_0|}{|\Psi_2|}
  &\sim
  \frac{(q^2+4)\,r_0^{\,2}}{8\,r^{2}}\sin^2\theta,
  \qquad
  \frac{|\Psi_1|}{|\Psi_2|}
  \sim
  \frac{(q^2+4)\,r_0^{\,2}}{16\,r^{2}}|\sin(2\theta)|.
\end{align}
}
Thus, at large distances the monopole contribution contained in $\Psi_2$ is dominant, whereas the $\mathcal{O}(r^{-5})$ terms represent progressively smaller multipole corrections.

\paragraph{Consistency with Petrov type and wormhole interpretation.}
Independently of the asymptotic analysis above, the algebraic type is determined by the curvature invariants: because $I^3\neq 27J^2$, the spacetime is algebraically general (Petrov type~I). The conditions $\Psi_0=\Psi_4$ and $\Psi_3=-\Psi_1$ do not indicate algebraic speciality, instead, they are compatible with a discrete symmetry that interchanges ingoing and outgoing null directions, as one would expect in a stationary wormhole spacetime with two equivalent asymptotic universes. In the limit $r\to \pm \infty$, the fact that $\Psi_2$ dominates shows that the Weyl tensor tends toward a Coulomb-type behavior, whereas the local structure near the throat remains algebraically general.

\paragraph{Minkowski limit.}
By setting the parameter \(q = 0\) in our NPW invariants, we observe that all invariants \(\Psi\) vanish, \(\Psi = 0\). Consequently, we recover the well-known Minkowski space-time expressed in Boyer–Lindquist coordinates, as previously discussed in Section \ref{section: gen sol}. Therefore, this solution is well behaved, and the Minkowski limit is obtained in the case where the \textit{efective mass monopole parameter} \(q\) is switched off.

\paragraph{Asymptotic Petrov character.}
Although the invariant condition $\mathcal{
I}^3 \neq 27 \mathcal{J}^2$ indicates that the spacetime is, in general, algebraically generic (Petrov type~I), the asymptotic behavior $\Psi_2 \sim \mathcal{O}(r^{-3})$ together with $\Psi_0,\Psi_1,\Psi_3,\Psi_4 \sim \mathcal{O}(r^{-5})$ implies that, in the limit $r \to \pm \infty$, the curvature invariants are dominated by the Coulomb-like contribution. In particular, one finds $\mathcal{I} = 3 \Psi_2^2 + \mathcal{O}(r^{-10})$, $\mathcal{J} = - \Psi_2^3 + \mathcal{O}(r^{-13})$.
Equivalently, by taking the asymptotic limits of $\mathcal{I}$ and $\mathcal{J}$, we recover Eq.~\eqref{Tipo D asimptotico}. This relation shows that the geometry becomes \emph{asymptotically} of Petrov type~D, even though it remains of type~I in the strong-field region ($r \ll \pm \infty$ ).

%%%%%%%%%%%%%%%%%%%%%%%%%%%%%
\section{Komar mass, Komar angular momentum and s-parameter}   %
%%%%%%%%%%%%%%%%%%%%%%%%%%%%%
\label{section: Komar cargas}

In this section, we compute the invariant conserved charges using the results presented in Appendix \ref{ApnediceCargasInv}. The mass $M$ is determined from the Komar integral associated with the stationary Killing vector $\partial_t$, following Komar’s original prescription \cite{Komar:1958wp,Nedkova:2011hx}. The angular momentum $J$ is obtained as the Komar charge corresponding to the axial Killing vector $\partial_\varphi$ \cite{Komar:1958wp,Nedkova:2011hx,Clement:2022pjr}.

Let $\xi=\partial_t$ and $\eta=\partial_\phi$ denote the stationary and axial Killing vector fields, respectively.
Their associated one-forms are given by $\xi^\flat = g(\xi,\cdot)=g_{tt}\,dt$ and $\eta^\flat = g(\eta,\cdot)=g_{\phi\phi}\,d\phi$.
Using these, we introduce the Komar mass and Komar angular momentum as follows:
\begin{equation}\label{eq:Komar_defs_trthph}
M_K=-\frac{1}{8\pi }\int_{S_r}\star d\xi^\flat,
\qquad
J_K=\frac{1}{16\pi }\int_{S_r}\star d\eta^\flat.
\end{equation}
The overall sign in \(M_K\) is chosen in the standard way so that, in the usual gauge, one has \(M_K > 0\) for the Schwarzschild solution.

Let us start by taking the exterior derivative of the 1-form $\eta^\flat$
\begin{equation*}
    d\eta^\flat = \partial_r\!\left( g_{\phi \phi} \right)\,dr\wedge d\phi + \partial_\theta\!\left( g_{\phi \phi} \right)\,d\theta\wedge d\phi.
\end{equation*}
We observe that there is no $t \, r$ component, i.e. $(d\eta^\flat)_{tr}=0$. Therefore, to compute the Komar angular momentum \eqref{eq:Komar_defs_trthph}, we apply \eqref{eq:Integral2FormaSobreSx}. However, using $H_{tr}=(d\eta^\flat)_{tr}$ in \eqref{eq:HodgeHMaestra}, it follows that, for this particular solution, 
\begin{equation}\label{eq:Komar_momentumAngular}
    J_K=0.
\end{equation}

On the other hand, we find that $d\xi^\flat=-df\wedge dt=\partial_r f\,dt\wedge dr+\partial_\theta f\,dt\wedge d\theta$. Hence, by \eqref{eq:HodgeHMaestra}, the only term that contributes to $H_{tr}$ is $(d\xi^\flat)_{tr}=\partial_r f$. Substituting in \eqref{eq:HodgeHMaestra}
\begin{equation*}
(\star d\xi^\flat)_{\theta\phi}
=
-\Delta_r\,\sin\theta\,\frac{\partial_r f}{f}
=
-\Delta_r\,\sin\theta\,\partial_r(\ln f),
\end{equation*}
introducing in \eqref{eq:Komar_defs_trthph}
\begin{align}
    M_K(r)
&=
-\frac{1}{8\pi }
\int_0^{2\pi}\!\!d\phi\int_0^\pi\!\!d\theta\;
(\star d\xi^\flat)_{\theta\phi}
\nonumber\\
&=
\frac{1}{8\pi }
\int_0^{2\pi}\!\!d\phi\int_0^\pi\!\!d\theta\;
\Delta_r\,\sin\theta\,\partial_r(\ln f)
\nonumber\\
&=\frac{1}{4}\int_0^\pi \Delta_r\,\sin\theta\left(q\frac{r_0}{\Delta_r}\right)d\theta
=\frac{q\,r_0}{4}\int_0^\pi \sin\theta\,d\theta \notag \\
&=\frac{q\,r_0}{2},
\end{align}
where $\ln f=q\Phi(r)+2s\ln\tan\frac{\theta}{2}$, then $\partial_r(\ln f)=q\,\Phi'(r),$ and $\Phi'(r)= \frac{r_0}{r^2+r_0^2} = \frac{r_0}{\Delta_r}$. Therefore the Komar mass at infinty is
\begin{equation}\label{eq:Komar_Masa}
    M_\infty=\lim_{x\rightarrow \infty} M_K(r)=\frac{q \, r_0}{2}.
\end{equation}

These results enable us to state with greater confidence that our spacetime is static, since there is no frame dragging and the azimuthal component of the flow vanishes, as shown in \eqref{eq:Komar_momentumAngular}. Furthermore, from \eqref{eq:Komar_Masa} it is clear that the parameter $q$ determines whether the mass is positive or negative, and this expression is independent of the parameter $s$. However, we must proceed with caution, because this parameter governs the asymptotic behaviour.

\paragraph{The disappearance of the NUT parameter.}
The twist or vorticity associated with $\xi$ is the 1-form
\begin{equation*}
\omega=\star\big(\xi^\flat\wedge d\xi^\flat\big)
\end{equation*}
which provides an entirely geometric characterization of how $\xi$ fails to be hypersurface-orthogonal.
In fact, by the Frobenius theorem, $\xi$ is hypersurface-orthogonal (equivalently, the spacetime is static in coordinates adapted to $\xi$) if and only if $\xi^\flat\wedge d\xi^\flat=0$, i.e., if and only if $\omega=0$.
For our diagonal metric one has $d\xi^\flat=-df\wedge dt$ and therefore $\xi^\flat\wedge d\xi^\flat=f\,dt\wedge df\wedge dt=0$, therefore $\omega=0$.

The NUT charge is defined as the monopole (Gauss) flux of the twist across a large 2-sphere,
$N\propto\int_{S_\infty}\omega$. Consequently, for our solution the NUT charge is zero:
\begin{equation}\label{eq:NUT_zero_conclusion}
 N=0.
\end{equation}
%................................................................................
\paragraph{Parameter $s$}
Let $S_x$ be the surface defined in Appendix \ref{ApnediceCargasInv}, and consider the induced 2-metric in coordinates $(\theta,\phi)$,
$d\sigma^2 = g_{\theta\theta}\,d\theta^2 + g_{\phi\phi}\,d\phi^2$. With $A=\Delta^\ell h$, a standard axis regularity test compares the azimuthal circumference $C = 2\pi\sqrt{g_{\phi\phi}}$ with the proper meridional distance $\rho = \int_0^\theta \sqrt{g_{\theta\theta}}\,d\theta'$. For a conical axis one finds $C = 2\pi\alpha\,\rho(\theta) + o(\rho)$ with finite $\alpha=
\lim_{\theta\to 0}\frac{C}{2\pi\rho}\in(0,\infty)$, leading to an angular deficit $\delta = 2\pi(1-\alpha)$ \cite{Vilenkin:1984ib,vandeMeent:2012gb}.

Now taking into account, for example, the north pole axis $\theta\to0$ 
\[
\sin\theta\sim\theta,
\qquad
\Delta_\theta\to r^2+r_0^2=\Delta_r,
\qquad
\left(\frac{\Delta_\theta}{\Delta_r}\right)^\ell\to 1
\]
therefore 
\[
A\sim
\Delta_r^{-s^2}\,e^{2qs\Phi(r)}\,\theta^{2s^2}
\quad\Longrightarrow\quad
A\xrightarrow[\theta\to0]{}0\ \text{ if }\ s\neq0.
\]
Consequently,
\[
g_{\theta\theta}\propto A(\theta)\sim \theta^{2s^2},
\qquad
g_{\phi\phi}\sim \sin^2\theta\sim \theta^2,
\]
which yields
\[
\rho\propto \int_0^\theta \theta'^{\,s^2}\,d\theta' \propto \theta^{1+s^2},
\qquad
C\propto \theta,
\]
\[
\therefore \frac{C}{2\pi\rho}\propto \theta^{-s^2}\to\infty\quad(s\neq0).
\]

Thus, for $s \neq 0$, the pole is not merely conical defect, instead, the parameter $s$ controls the axial/polar geometry. Consequently, $s \neq 0$ generates an \emph{axis singularity that is stronger than a conical defect}, meaning it cannot be interpreted as a simple cosmic-string-like deficit angle.

\subsection{Ring conical excess and axial conical defects}

It is useful to distinguish two different notions of conical singularity in the present geometry. The conical singularity discussed by Gibbons and Volkov in \cite{Gibbons:2016bok} is localized at the ring.
By contrast, the conical test discussed above concerns the polar axis. These are geometrically different loci.

Let us first consider the asymptotically flat sector, \(s=0\), where the metric reduces to the Gibbons-Volkov-type ring-wormhole geometry. The explicit and spatial part of the metric is
\(
d\ell_{\rm sp}^{2}
=
e^{-q\Phi}
\left[
\Delta^\ell
\left(
dr^2+\Delta_r\,d\theta^2
\right)
+
\Delta_r\sin^2\theta\,d\varphi^2
\right]
\), where \(\ell=1+\frac{q^2}{4}\). The ring is located at \(r=0,\quad\theta=\frac{\pi}{2}\). Near this ring, we introduce the local coordinate \(y=\mathtt r_0\cos\theta\), in the limit \(r\simeq 0\), \(\theta\simeq \pi/2\), one has 
{\footnotesize
\[
\Delta_r\simeq \mathtt r_0^2,
\quad
\Delta_r\,d\theta^2\simeq dy^2,
\quad
\Delta\simeq \frac{r^2+y^2}{\mathtt r_0^2},
\quad
e^{-q\Phi}\simeq 1,
\quad
\sin\theta\simeq 1 .
\]
}
Therefore, the local spatial geometry near the ring becomes is
\[
d\ell_{\rm sp}^{2}
\simeq
\left(
\frac{r^2+y^2}{\mathtt r_0^2}
\right)^\ell
\left(
dr^2+dy^2
\right)
+
\mathtt r_0^2\,d\varphi^2 .
\]
We now introduce polar coordinates in the two-dimensional plane transverse to the ring, given by \(r=\mathtt r_0\chi\cos\alpha, \quad y=\mathtt r_0\chi\sin\alpha \), then \(dr^2+dy^2=\mathtt r_0^2\left(d\chi^2+\chi^2\,d\alpha^2\right)\), and the local metric takes the form
\[
d\ell_{\rm sp}^{2}
\simeq
r_0^2\chi^{2\ell}
\left(
d\chi^2+\chi^2\,d\alpha^2
\right)
+
r_0^2\,d\varphi^2 .
\]
The transverse two-metric can be brought to a locally flat form by defining
\[
\mathtt R=\frac{\chi^{\ell+1}}{\ell+1},
\qquad
\vartheta=(\ell+1)\alpha .
\]
Indeed, this gives \( d\ell_{\rm sp}^{2} \simeq \mathtt r_0^2 \left( d\mathtt R^2 + \mathtt R^2\,d\vartheta^2 \right) + \mathtt r_0^2\,d\varphi^2 \), however, since the angular coordinate \(\alpha\) has period \(2\pi\), the angular coordinate \(\vartheta\) has period \(\Delta\vartheta_{\rm period}=2\pi(\ell+1)\), therefore, the angular defect with respect to the standard value \(2\pi\) is
{\footnotesize
\[
\delta_{\rm ring}
=
2\pi-\Delta\vartheta_{\rm period}
=
2\pi-2\pi(\ell+1)
=
-2\pi\ell
=
-2\pi
\left(
1+\frac{q^2}{4}
\right).
\]
}
This is a negative deficit, or equivalently, a conical excess localized at the ring. With the identification \(q=2\sigma\), this becomes \(\delta_{\rm ring}=-2\pi(1+\sigma^2)\), which is precisely the conical excess found by Gibbons and Volkov in \cite{Gibbons:2016bok}. In their interpretation, this corresponds to a circular source with negative tension.

Thus, there is no inconsistency with the Gibbons–Volkov analysis. The extended \(s=0\) geometry exhibits the same ring-like conical excess as in Ref. \cite{Gibbons:2016bok}. The axial computation merely demonstrates that the \(s\)-dependent polar singularity does not correspond to an ordinary conical defect.

Finally, in the cut-and-paste construction used to define the wormhole manifold, the central region containing the ring is removed. The physical domain is
\begin{equation*}
r\in
\left(
-\infty,-\frac{q r_0}{2}
\right]
\cup
\left[
\frac{q r_0}{2},\infty
\right),
\end{equation*}
and the two boundaries \(r=\pm q r_0/2\) are identified with each other. Consequently, although the extended geometry possesses the same ring conical excess as the Gibbons-Volkov solution, this defect is not part of the final cut-and-paste wormhole manifold. Therefore, the physical wormhole spacetime is protected from the ring conical defect by the throat construction itself.

%%%%%%%%%%%%%%%%%%%%%%%%%%%%%%%%%%%%%%%%%%%%%%%%%%%%%%%%%%%%%%%%%%%%%
\paragraph{Relation with scalar and phantom-field constructions.}   %
%%%%%%%%%%%%%%%%%%%%%%%%%%%%%%%%%%%%%%%%%%%%%%%%%%%%%%%%%%%%%%%%%%%%%

A key question is how the present vacuum geometry relates to the scalar and phantom-field models of Gibbons and Volkov. Their construction starts from Einstein gravity with a scalar field (including the phantom sector) and uses duality rotations and analytic continuations to obtain ring-wormhole geometries. The resulting ring wormhole is nonetheless a vacuum spacetime. This is especially relevant for \(s=0\), where our solution reduces to the same ring-wormhole geometry of Ref. \cite{Gibbons:2016bok}, including the local conical excess at the ring. Thus, the link to scalar or phantom fields is only indirect: the final geometry is vacuum, and the scalar field appears only in the solution-generating procedure.

For \(s\neq0\) the situation changes. The parameter \(s\) alters the axial structure via
\[
f=e^{q\Phi}\tan^{2s}\frac{\theta}{2},
\qquad
h=\Xi^{s^2}e^{2qs\Theta},
\]
yielding a non-asymptotically flat geometry with a stronger axial singularity. Still, the spacetime is Ricci-flat and contains no scalar or phantom stress-energy. Therefore, \(s\) is not a scalar charge, but a purely geometric deformation parameter that modifies the global structure of the vacuum solution.

%%%%%%%%%%%%%%%%%%%%%%%%%%%%%%%%%%%%%%%%%%%%%%%%%%%%%
\section{Trapped and marginally trapped surfaces}   %
%%%%%%%%%%%%%%%%%%%%%%%%%%%%%%%%%%%%%%%%%%%%%%%%%%%%%
\label{section: trapped surfaces}

For our porpouse, we will employ a geometrically preferred NP tetrad constructed from the static Killing field, first we select the unit future-directed 4-velocity field of static observers, given by:

\begin{equation}  \label{4VelocidadEstatico}
  u^\mu := \frac{\xi^\mu}{\sqrt{-\xi^2}},
  \qquad |
  \qquad
  u^\mu u_\mu = -1,
\end{equation}
where $\xi=\partial_t$ and $\xi ^2= \xi ^{\mu} \xi_{\mu}$, and, by selecting the orthogonal proper acceleration $u^\nu \nabla_\nu u_\mu$ and subsequently normalizing it, we obtain
\begin{equation}\label{aceleracion normalizada}
    s^\mu = \frac{a^\mu}{\sqrt{a^\mu a_\mu}},
\end{equation}
fulfilling $a^{\mu}u_{\mu}=0$ and $a^{\mu}a_{\mu}=1$.

The second step is to construct two vectors that are orthogonal to both $u^\mu$ and $a^\mu$. Making use of the axisymmetry of the space-time, we may select one of them as $b^\mu = g^{\phi \phi}\partial_\phi^\mu$. This choice guarantees that $b^{\mu} b_\mu = 0$ and that $b^{\mu} u_{\mu} = b^{\mu} a_{\mu} = 0$. The other vector is defined by $\widehat{e}^\mu = g^{\mu \sigma} \varepsilon_{\sigma \alpha\beta\gamma}\,u^\alpha s^\beta b^\gamma $ ,where $\varepsilon_{\sigma \alpha\beta\gamma}$ denotes the Levi-Civita symbol. After normalizing this vector, we obtain:
\begin{equation}\label{aceleracion normalizada}
    e^\mu = \frac{\widehat{e}^\mu}{\sqrt{\widehat{e}^\mu \widehat{e}_\mu}},
\end{equation}
By construction, $e^\mu e_\mu = 1$, and $e^{\mu} u_{\mu} = e^{\mu} s_{\mu} = e^{\mu} b_{\mu} = 0$.

Now, we construct the NP null tetrad as
\begin{align}
    &l^\mu := \frac{1}{\sqrt{2}}\bigl(u^\mu + s^\mu\bigr),
  \qquad
  n^\mu := \frac{1}{\sqrt{2}}\bigl(u^\mu - s^\mu\bigr), \notag \\
  &m^\mu := \frac{1}{\sqrt{2}}\bigl(e^\mu + i\,b^\mu\bigr),
  \qquad
  \bar m^\mu := \frac{1}{\sqrt{2}}\bigl(e^\mu - i\,b^\mu\bigr).
\end{align}
These satisfy the NP inner-product relations
\begin{align*}
  &l^\mu l_\mu = n^\mu n_\mu = m^\mu m_\mu = \bar m^\mu \bar m_\mu = 0,
  \qquad \\
  & \qquad l^\mu n_\mu = -1,
  \qquad
  m^\mu \bar m_\mu = +1,
\end{align*}
with all other contractions vanishing $l\cdot m=l\cdot\bar m=n\cdot m=n\cdot\bar m=0$. The construction of the NP geometrically preferred tetrad was carried out following the methodologies outlined in Refs.~\cite{Gourgoulhon:2005ng,Maartens:1997fg,Adamo:2009vu,Ashtekar:2004cn}.

For the sake of a more concise and compact notation, we introduce the following definitions. Let
\[
z_N{}^\mu=\{\,l^\mu,\; n^\mu,\; m^\mu,\; \bar m^\mu\,\}
\]
denote the geometrically preferred null Newman-Penrose tetrad, with tetrad indices
\(M,N,P\in\{l,n,m,\bar m\}\).
We introduce the connection coefficients of the Levi-Civita connection
in the null frame by
\begin{equation}\label{CoefSpinCompactos}
Z_{MNP}
\;:=\;
z_M{}^\mu\, z_N{}^\nu\, \nabla_\mu z_{P\nu},
\qquad
Z_{NMP}=-Z_{NMP}.
\end{equation}
The antisymmetry in the final two indices arises directly from metric compatibility, \(\nabla_\mu g_{\alpha\beta}=0\), when this condition is applied to the inner products of the tetrad vectors. The index \(M\) indicates the direction along which the derivative is evaluated, whereas \(N\) and \(P\) denote the covariant components of the tetrad.

In this notation, the Newman-Penrose spin coefficients are identified as the
following components of \(Z_{MNP}\):
\begin{equation}\label{CoefSpin}
\begin{aligned}
\kappa &= Z_{l m l}, &
\sigma &= Z_{m m l}, \\
\rho   &= Z_{\bar m m l}, &
\tau   &= Z_{n m l}, \\
\pi    &= - Z_{l \bar m n}, &
\mu    &= Z_{\bar m m n}, \\
\lambda&= Z_{\bar m \bar m n}, &
\nu    &= Z_{n \bar m n}, \\
\epsilon &= \tfrac12\!\left(Z_{l n l}-Z_{l \bar m m}\right), &
\gamma   &= \tfrac12\!\left(Z_{n l n}-Z_{n m \bar m}\right), \\[4pt]
\alpha   &= \tfrac12\!\left(Z_{\bar m n l}-Z_{\bar m \bar m m}\right), &
\beta    &= \tfrac12\!\left(Z_{m l n}-Z_{m m \bar m}\right).
\end{aligned}
\end{equation}

In our situation, we find
{\small
\begin{subequations}\label{CoefSpinEvaluados}
\begin{align}
    \kappa & =\pi=-\tau=-\nu= -\frac{ \ell r_0^2 \sqrt{f} \sin \theta \cos \theta }{ 2 \sqrt{2} \sqrt{ \Delta^\ell \Delta_r } \Delta_\theta }, \\
    \sigma &=\lambda= -\frac{ \ell r_0^2 \sqrt{f} r \sin ^2 \theta }{ 2 \sqrt{2} \sqrt{\Delta^\ell} \Delta_\theta \Delta_r }, \\
    \rho   &=\mu = \frac{\sqrt{f}}{ 2 \sqrt{2} \sqrt{\Delta^\ell} \Delta_\theta \Delta_r } \bigg( r_0^2 \cos^2 \theta ( ( \ell - 2 ) r + q r_0)\notag  \\
    &\qquad \qquad -r ( \ell r_0^2 - q r_0 r + 2 r^2)\bigg) , \label{CoefSpinRhoMu}\\
    \epsilon &= \gamma  =-\frac{ q r_0 \sqrt{f} }{4 \sqrt{2} \sqrt{ \Delta^\ell } \Delta_r }, \\
    \alpha   &= -\beta = \frac{\sqrt{f}}{2 \sqrt{2} \, \tan{\theta} \, \sqrt{ \Delta^\ell \Delta _r } }.
\end{align}
\end{subequations}
}

By analyzing \eqref{CoefSpinEvaluados}, we observe that in all cases the spin coefficients are purely real, in other words, $\Im(Z_{MNP}) = 0$. Adopting the purely geometric interpretation of these coefficients, the complex optical scalars associated with the null congruences generated by $l^\mu$ and $n^\mu$ are encoded in \eqref{CoefSpinRhoMu} in the following manner:
\begin{align*}
    \rho &= \underbrace{\Re(\rho)}_{\text{expansion of }l}
         + i\,\underbrace{\Im(\rho)}_{\text{twist of }l}, \\
    \mu &= \underbrace{\Re(\mu)}_{\text{expansion of }n}
        + i\,\underbrace{\Im(\mu)}_{\text{twist of }n}.
\end{align*}
Consequently, \emph{the null congruences are non-twisting (i.e., hypersurface-orthogonal)}. This aligns with the static nature of the spacetime. In a static region there is no frame dragging, and one can select the transverse frame so that no spurious phase rotation appears in $m^\mu$.

\paragraph{Important caveat (geodesic/affine conditions).}
Even though our adapted null normals are twist-free, they are \emph{not required} to be geodesic or affinely parametrised. Specifically, the conditions $\kappa\neq 0$ and/or $\epsilon+\bar\epsilon\neq 0$ signal that $l^\mu$ is not a geodesic generator and/or is not affinely parametrized. 

\paragraph{2-surfaces and induced metric.}\label{Comienzo SupAtrapadas Throat con NP}
We now employ this decomposition to examine trapped and marginally trapped surfaces, as well as potential wormhole throats. Let \(S\) be a spacelike 2-surface in the space-time manifold. Given future-directed null normal vectors \(l^\mu\) and \(n^\mu\), orthogonal to \(S\) and normalized such that \(l \cdot n = -1\), the induced metric (i.e., the projection tensor) on the tangent bundle \(TS\) is given by
\begin{equation*}
  q_{\mu\nu} := g_{\mu\nu} + l_\mu n_\nu + n_\mu l_\nu,
  \quad | \quad 
  q^\mu{}_\nu l^\nu = q^\mu{}_\nu n^\nu = 0.
\end{equation*}

\paragraph{Null expansions (covariant definition).}
The outgoing/ingoing null expansions are defined quasi-locally by
\begin{equation*}
  \theta_{(l)} = q^{\mu\nu}\nabla_\mu l_\nu,
  \qquad
  \theta_{(n)} = q^{\mu\nu}\nabla_\mu n_\nu.
\end{equation*}
These definitions are purely geometric and do \emph{not} require $l^\mu$ or $n^\mu$ to be geodesic.

\paragraph{Relation to NP coefficients.}
If $m^\mu$ and $\bar m^\mu$ span the tangent space of $S$, then $q^{\mu\nu} = 2\, m^{(\mu}\bar m^{\nu)}$.
In this case, the expansions are directly related to the NP optical scalars:
\begin{align}
  \theta_{(l)} &= -\bigl(\rho+\bar\rho\bigr) = -2\,\Re(\rho), \\
  \theta_{(n)} &= -\bigl(\mu+\bar\mu\bigr) = -2\,\Re(\mu).
\end{align}
In our solution $\rho,\mu\in\mathbb{R}$, hence $\theta_{(l)}=\theta_{(n)}=-2\rho$.

\paragraph{Trapped and marginally trapped surfaces.}\label{Final SupAtrapadas Throat con NP}
A spacelike 2-surface $S$ is said to be (future) trapped if
\begin{equation}  \label{eq:trapped_def}
  \theta_{(l)}<0 \quad \text{and}\quad \theta_{(n)}<0,
\end{equation}
everywhere on $S$. A marginally outer trapped surface (MOTS) satisfies
\begin{equation}\label{eq:MOTS_def}
  \theta_{(l)}=0,\qquad \theta_{(n)}<0,
\end{equation}
with the external condition determined by the selected orientation, that is, by which null normal is designated as outgoing. In our case, only the condition specified in \eqref{eq:trapped_def} can be satisfied.
Taking into account $r>0$ and $q>0$, we found that the region where $\theta_{(l)}=-2 \rho<0$ appears when $0 < \frac{r}{r_0} < \frac{q }{2}$.

%\textcolor{red}{All the formalism developed from \ref{Comienzo SupAtrapadas Throat con NP} to \ref{Final SupAtrapadas Throat con NP} is formulated on the basis of the foundational works} \cite{Newman:1961qr,Hayward:1994yy,Hochberg:1997wp,Hochberg:1998ha,Hochberg:1998ii}.
All the formalism developed in this section is formulated on the basis of the foundational works \cite{Newman:1961qr, Hayward:1994yy, Hochberg:1997wp, Hochberg:1998ha, Hochberg:1998ii}.

%%%%%%%%%%%%%%%%%%%%%
\section{Wormholes} %
%%%%%%%%%%%%%%%%%%%%%
\label{section: throat}

In this section, we will construct wormholes using the solution \eqref{metric} with $s = 0$.

Let
\begin{equation}
\label{perp metric}
    \hat{g}_\perp
    = \Delta^\ell d\theta^2
    + \sin^2 \theta d\phi^2 .
\end{equation}
Here, $\Delta = 1 - \epsilon_0^2 \sin^2 \theta$ and $\epsilon_0 = \tfrac{r_0}{\sqrt{\Delta_r}} \in ( 0, 1 ]$ with $r = const$.
It has the Killing vector $\partial_\phi$.
If $r \to \pm\infty$, then $\epsilon_0 \to 0$, so that $\hat{g}_\perp \to d\theta^2 + \sin^2 \theta d\phi^2$, which is the metric of a 2-sphere.
%It is well known that the 2-sphere is a surface of revolution. % parametrized by $x = \cos \phi \sin \theta$, $y = \sin \phi \sin \theta$, $z = \cos \theta$.
We assume that the metric \eqref{perp metric} corresponds to a surface of revolution embedded in the Euclidean space $\mathbb{R}^3$ parametrized by
$( \phi, \theta ) \mapsto ( \rho(\theta), \phi, z (\theta) )$.
Then,
\begin{equation}
\label{perp rev surface}
    \hat{g}_\perp = dz^2 + d\rho^2 + \rho^2 d\phi^2 = ( z_{, \theta}^2 + \rho_{, \theta}^2 ) d\theta^2 + \rho^2 d\phi^2
\end{equation}
Comparing Eqs. \eqref{perp metric} and \eqref{perp rev surface}, we have $\rho = \sin \theta$ and $z_{, \theta}^2 = \Delta^\ell - \cos^2 \theta$.
Using $\Delta^\ell \geq 1 - \ell \epsilon_0^2 \sin^2 \theta$, we get $z_{, \theta}^2 \geq ( 1 - \ell \epsilon_0^2 ) \sin^2 \theta$.
If $\epsilon_0 \leq \tfrac{1}{\sqrt{\ell}}$, then $z_{, \theta}^2 \geq 0$.

Let
\begin{equation}
    \mathscr{A}_\perp ( \epsilon_0 )
    = 2 \pi \int_0^\pi \sqrt{ \Delta^\ell } \sin \theta d\theta
\end{equation}
be the area given by the metric \eqref{perp metric}.
Let $\Delta_\ell = \Delta ( \epsilon_0 = \tfrac{1}{\sqrt\ell} )$.
Given that $\Delta \geq \Delta_\ell$, then $\mathscr{A}_\perp ( \epsilon_0 ) \geq \mathscr{A}_\perp ( \tfrac{1}{\sqrt\ell} )$.
Hence, $\mathscr{A}_\perp$ has a minimum at $\epsilon_0 = \tfrac{1}{\sqrt\ell}$.
The surfaces with minimal area are given by the solution of $z_{, \theta}^2 = \Delta_\ell^\ell - \cos^2 \theta$ with the initial condition $z ( \theta = \tfrac{\pi}{2} ) = 0$.

\begin{figure}
    \centering
    \begin{subfigure}{0.45\textwidth}
        \includegraphics[width=1\linewidth]{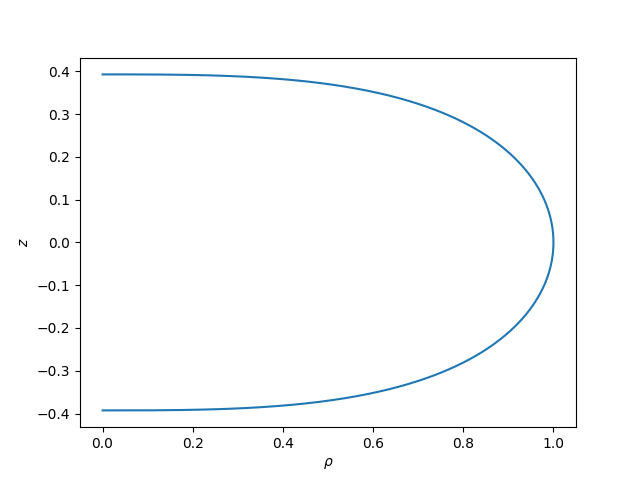}
        \caption{Shape curve.}
        \label{fig: shape_2}
    \end{subfigure}
\vfill
    \begin{subfigure}{0.45\textwidth}
        \includegraphics[width=1\linewidth]{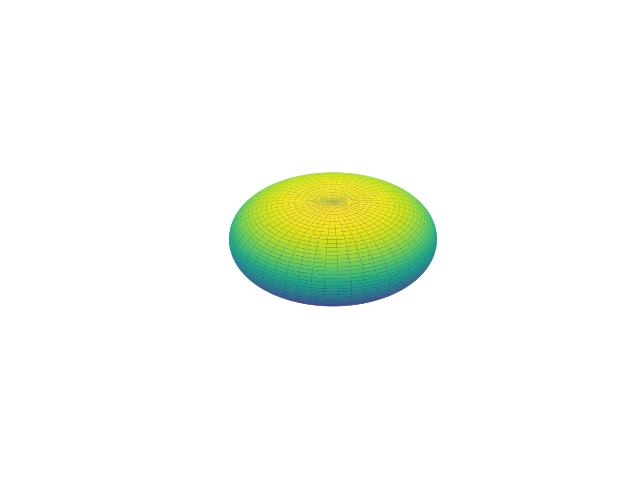}
        \caption{Surface of revolution.}
        \label{fig: surface_2}
    \end{subfigure}
\caption{Shape curve and its corresponding surface of revolution with $\ell = 2$.}
\label{fig: ell_2}
\end{figure}

\begin{figure}
    \centering
    \begin{subfigure}{0.45\textwidth}
        \includegraphics[width=1\linewidth]{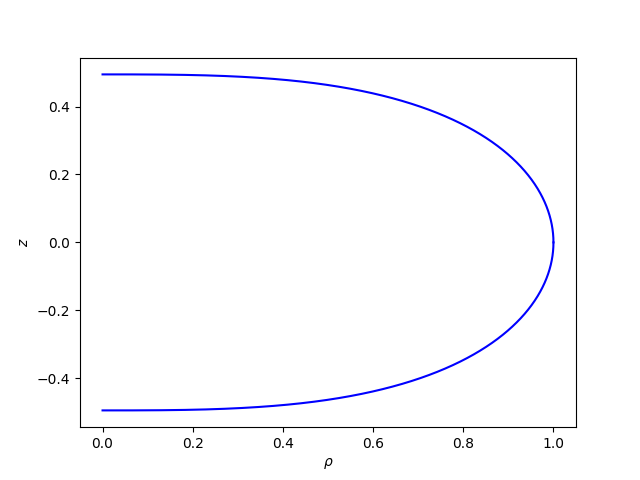}
        \caption{Shape curve.}
        \label{fig: shape_oo}
    \end{subfigure}
\vfill
    \begin{subfigure}{0.45\textwidth}
        \includegraphics[width=1\linewidth]{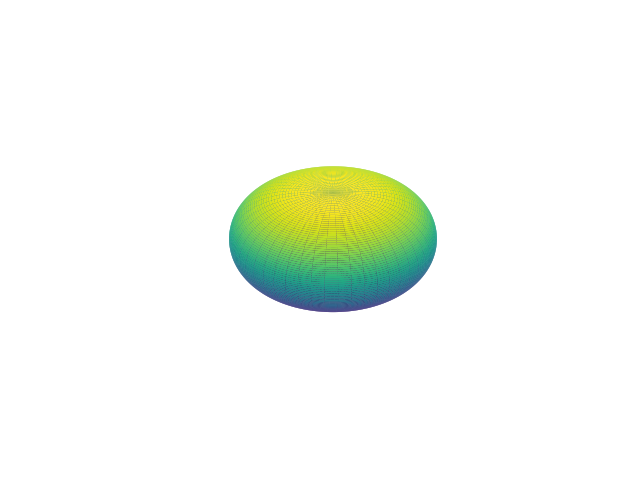}
        \caption{Surface of revolution.}
        \label{fig: surface_oo}
    \end{subfigure}
\caption{Shape curve and its corresponding surface of revolution with $\ell \to \infty$.}
\label{fig: ell_oo}
\end{figure}

\begin{figure}
    \centering
    \includegraphics[width=1.0\linewidth]{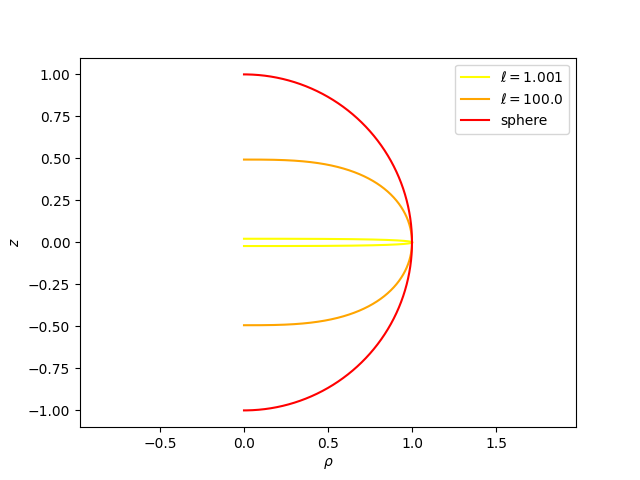}
    \caption{Shape curves for the sphere and for surfaces with $\ell = 1.001$ and $\ell = 100.0$.}
    \label{fig: shapes}
\end{figure}

If $\ell = 1$, then $z = 0$ on $I_\theta$, so that the metric \eqref{perp metric} corresponds to a disk of unit radius.
This result agrees with that obtained by Gibbons and Volkov in \cite{Gibbons:2016bok}.
When $\ell = 2$, the surface is parametrized by
\begin{equation}
\label{surface rev q2}
    x = \cos \phi \sin \theta ,\,
    y = \sin \phi \sin \theta ,\,
    8 z = 2 \theta - \sin 2 \theta - \pi .
\end{equation}
The image of $z$ on $I_\theta$ is $[ -\tfrac{\pi}{8}, \tfrac{\pi}{8} ]$.
In Figure \ref{fig: ell_2}, we plot the shape curve and its corresponding surface of revolution for $\ell = 2$.
Given that $\Delta_\ell^\ell \to \exp( -\sin^2 \theta )$ as $\ell \to \infty$, we can define the superface of revolution given by $z_{, \theta}^2 = \exp( -\sin^2 \theta ) - \cos^2 \theta$.
Observe in Figure \ref{fig: shape_oo} that $z$ approaches $0.5$ at $\theta = 0$ and $-0.5$ at $\theta = \pi$.
By rotating this curve, the surface shown in Figure \ref{fig: surface_oo} is created.
In Figure \ref{fig: shapes}, we compare the shape curves for the sphere and for the surfaces with $\ell = 1.001$ and $\ell = 100.0$.
The shape curve tends to a line as $\ell$ approaches 1, while it tends to curve in Figure \ref{fig: shape_oo} for larger values of $\ell$.

The role of the parameter \(\ell\) can be understood from its relation with \(q\),
\(
\ell=1+\frac{q^2}{4}.
\)
Thus, the case \(\ell=1\) corresponds to the degenerate limit \(q=0\). Since the Komar mass is proportional to \(q\), see Eq. \eqref{eq:Komar_Masa}, this limit corresponds to switching off the effective mass parameter. Geometrically, the minimal embedding profile satisfies \(z(\theta)=0\) for the whole angular interval. Hence the throat reduces to a disk of unit radius, as in the Gibbons-Volkov limit.

For \(\ell>1\), the parameter \(q\) is nonzero and the throat is lifted away from the plane \(z=0\). The embedding then becomes a closed surface of revolution. This surface should not be interpreted as a round two-sphere, it is better described as an oblate-like surface, with fixed equatorial radius and an \(\ell\)-dependent finite height. Increasing \(\ell\) deforms the throat away from the disk-like degenerate geometry and toward the limiting profile obtained when \(\ell\to\infty\). This interpolation is displayed in Figure \ref{fig: shapes}, where the profile approaches a straight line as \(\ell\to1\), while for large \(\ell\) it approaches the limiting curve shown in Figure \ref{fig: ell_oo}.

Finally, we stress that the coordinate \(z\) used in these diagrams is only an auxiliary Euclidean embedding coordinate.

Let  $\mathscr{A} = \tfrac{\Delta_r}{f} \mathscr{A}_\perp$ be the area of the 2-hypersurface defined by constant $t$ and $r$.
Inequality $\epsilon_0 \leq \tfrac{1}{\sqrt{\ell}}$ implies $\tfrac{r}{r_0} \geq \vert \tfrac{q}{2} \vert$ or $\tfrac{r}{r_0} \leq -\vert \tfrac{q}{2} \vert$.
Given that $\tfrac{\Delta_r}{f}$ and $\mathscr{A}_\perp$ has a minimum at $r = \tfrac{q}{2} r_0$, $\mathscr{A}$ also has a minimum at $r = \tfrac{q}{2} r_0$.

Before introducing the cut-and-paste construction, let us clarify its role. The extended \(s=0\) geometry already admits a ring-wormhole interpretation: the limits \(r\to+\infty\) and \(r\to-\infty\) correspond to two asymptotic regions connected through the disk bounded by the ring. Therefore, the cut-and-paste procedure is not needed in order to produce two asymptotic ends.

Its purpose is instead to define a regularized physical manifold in which the pathological ring region is removed. The extended geometry contains a ring at \(r=0,\quad\theta=\frac{\pi}{2}\), where the curvature singularity and the conical ring defect are localized. In the cut-and-paste construction, the central region containing this ring is excised, and the two exterior boundaries are identified. This replaces the original ring-defect crossing by a thin-shell throat located at \(r_G=\frac{q r_0}{2}\). The price of this regularization is the appearance of a surface stress-energy tensor on the throat. As shown by the Israel-junction analysis, the corresponding thin shell violates the null energy condition on the surface.

We consider a simple cut-paste construction of a wormhole \cite{LorentzianWormholes} for $q \neq 0$.
Let $\mathfrak{U}$ be a spacetime endowed with the metric \eqref{metric} with $s = 0$ and $r_0 > 0$.
%Now we take two different spacetimes $\mathfrak{U}$, and remove from them the 4-dimensional regions described by $U_+ = \{ r > \tfrac{q}{2} r_0 : q > 0 \}$ and $U_- = \{ r < -\tfrac{q}{2} r_0 : q > 0 \}$.
Let $U_+ = \{ r > \tfrac{q}{2} r_0 : q > 0 \}$ and $U_- = \{ r < -\tfrac{q}{2} r_0 : q > 0 \}$ be two manifolds arising from two different spacetimes $\mathfrak{U}$.
Let $\partial U_+ = \{ r = \tfrac{q}{2} r_0 \}$ and $\partial U_- = \{ r = -\tfrac{q}{2} r_0 \}$ be the boundaries of $U_+$ and $U_-$, respectively.
These manifolds are glued together by identifying their boundaries, $\partial U_+ = \partial U_- = \partial U$.
The resulting spacetime is $U$.
The functions $\Delta$ and $\Delta_r$ are the same for $U$; however, $f$ is given by $f ( r ) = \exp( q \arctan(\vert \tfrac{r}{r_0} \vert) ) $, with $q > 0$.
Note that the metric components are continuous on $\partial U$ and the domain of $r$ is $I_r = ( -\infty, -\tfrac{q}{2} r_0 ] \cup [ \tfrac{q}{2} r_0, \infty )$.
The throat of the wormhole is at $\partial U$.
In the remaining sections, we consider the spacetime $U$.

%%%%%%%%%%%%%%%%%%%%%%%%%%%%%%%%%%%%%
\section{The embedding surfaces}    %
%%%%%%%%%%%%%%%%%%%%%%%%%%%%%%%%%%%%%
\label{section: embedding surfaces}

In this section, we will investigate the embedding surface.

Let
\begin{equation}
\label{2D metric}
    \hat{g}^{(2)}
    = \frac{1}{f}\left[
        \Delta^\ell dr^2 
        + \varepsilon_0^2 \Delta_r d\phi^2\right]
     ,
\end{equation}
be the 2-dimensional hypersurface defined by constant $t$ and $\theta \in ( 0, \pi )$.
Here, $\Delta = 1 - \varepsilon_0^2 \tfrac{r_0^2}{\Delta_r}$ and $\varepsilon_0 = \sin \theta \in ( 0, 1 ]$.
This hypersurface is embedded in a 3-dimensional Euclidean space, which is parametrized by cylindrical coordinates,
\begin{equation}
\label{embedding surface}
     \hat{g}^{(2)}
    = dz^2 + d\rho^2 + \rho^2 d\phi^2
    = ( z_{, r}^2 + \rho_{, r}^2 ) dr^2 + \rho^2 d\phi^2 .
\end{equation}
Comparing Eqs. \eqref{2D metric} and \eqref{embedding surface}, we find $\rho = \sqrt{\tfrac{\Delta_r}{f}} \varepsilon_0$ and $z_{, r}^2 = \frac{
        \Delta^\ell \Delta_r
        - \varepsilon_0^2 ( \vert r \vert - \tfrac{q}{2} r_0 )^2
    }{ f \Delta_r }$.

Using the inequality $\Delta^\ell > 1 - \ell \varepsilon_0^2 \tfrac{r_0^2}{\Delta_r}$, we get
\begin{equation}
\label{ineq emb surface}
    \Delta_r f z_{, r}^2
    > ( 1 - \varepsilon_0^2 ) r^2
    + q \varepsilon_0^2 r_0 \vert r \vert
    + ( 1 - \varepsilon_0^2 - \tfrac{q^2}{2} \varepsilon_0^2 ) r_0^2 .
\end{equation}
Setting $\varepsilon_0 = 1$ into Ineq. \eqref{ineq emb surface}, it reduces to
\begin{equation}
    z_{, r}^2 > \tfrac{ q r_0^2 }{ \Delta_r f } ( \vert \tfrac{r}{r_0} \vert  - \tfrac{q}{2} )  \geq 0
\end{equation}
for all $r \in I_r$.
If $\varepsilon_0 < 1$, we rewrite Ineq. \eqref{ineq emb surface} as
\begin{equation}
    z_{, r}^2
    >  \tfrac{r_0^2 ( 1 - \varepsilon_0^2 )}{f \Delta_r} (
        (
            \vert \tfrac{r}{r_0} \vert
            + \tfrac{q}{2} \tfrac{1 + \varepsilon_0^2}{1 - \varepsilon_0^2}
        ) ( 
            \vert \tfrac{r}{r_0} \vert
            - \tfrac{q}{2}
        )
        + \ell 
    )> 0
\end{equation}
for all $r \in I_r$.

\begin{figure}
    \centering
    \begin{subfigure}{0.45\textwidth}
        \includegraphics[width=1\linewidth]{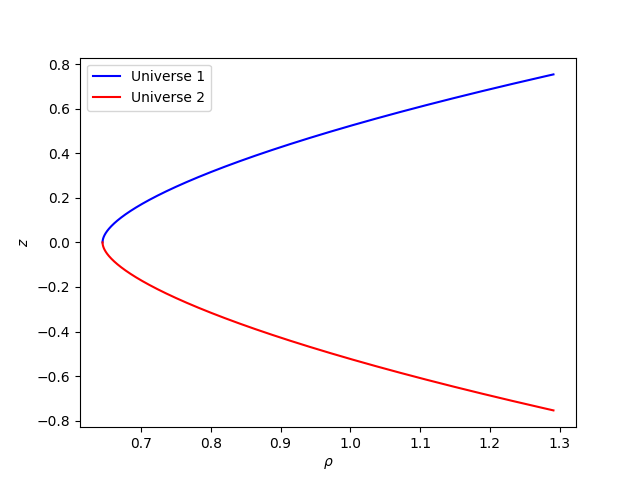}
        \caption{Profile curve}
        \label{fig: profile curve}
    \end{subfigure}
\vfill
    \begin{subfigure}{0.45\textwidth}
        \includegraphics[width=1\linewidth]{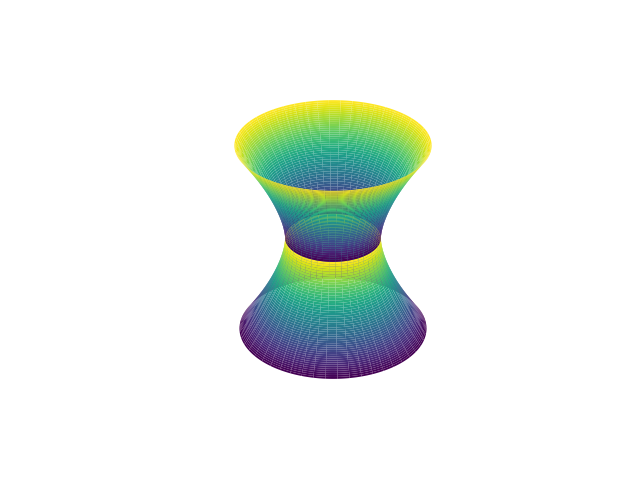}
        \caption{Embedding surface.}
        \label{fig: embbeding surface}
    \end{subfigure}
\caption{Profile curve and embedding surface of the wormhole with $q = 2$, $r_0 = 1$ and $\theta = \tfrac{\pi}{2}$.}
\label{fig: throat}
\end{figure}

In Figure \ref{fig: profile curve}, we plot the profile curve with $q = 2$, $r_0 = 1$ and $\theta = \tfrac{\pi}{2}$.
Its corresponding surface of revolution is shown in Figure \ref{fig: embbeding surface}.
Observe that the graph of profile curve is symmetric with respect to the $\rho$-axis, which implies that the surface is symmetric with respect to the plane $z = 0$.

%%%%%%%%%%%%%%%%%%%%%
\section{Geodesics} %
%%%%%%%%%%%%%%%%%%%%%
\label{section: geodesics}

In \cite{Gibbons:2016bok}, the conditions for the existence of solutions to the equations of motion with $\theta = 0$ or $\theta = \pi$ are found.
The geodesics along the symmetry axis traverse the wormhole.
In this section, we will determine the conditions for constant $\theta \in ( 0, \pi )$.

The equations of motion are obtained from the Lagrangian
\begin{equation}
    \mathscr{L}
    = \frac{
        \Delta^\ell (
            \dot{r}^2
            + \Delta_r \dot{\theta}^2
        )
        + \Delta_r \sin^2 \theta \dot{\phi}^2
    }{2 f}
    - \frac{f}{2}\, \dot{t}^2 ,
\end{equation}
where the dot denotes the derivative with respect to the affine parameter $\tau$.
Since $\mathscr{L}$ does not depend on $t$ and $\phi$, then $\tfrac{ \partial \mathscr{L} }{ \partial \dot{t} } = -E$ and $\tfrac{ \partial \mathscr{L} }{ \partial \dot{\phi} } = J$, so that
\begin{align}
    \dot{t}
&   = \tfrac{E}{f} ,
\\\label{eq motion phi}
    \dot{\phi}
&   = \tfrac{ J f }{ \varepsilon_0^2 \Delta_r }     ,
\end{align}
where $E$ and $J$ are  constants of integration. %, and $\varepsilon_0 = \sin \theta_0 \in ( 0, 1 ]$.
$E$ can be interpreted as the energy of the system and $J$ as the angular momentum.
To avoid velocities that exceed the speed of light, we impose $\dot{t} > 0$.
Thus, $E > 0$.

Combining Eq. \eqref{eq motion phi} and the equation of motion for $r$,
\begin{equation}
\label{eq motion r}
    \Delta^\ell \dot{r}^2
    =  E
    - \tfrac{ J^2 f^2 }{ \varepsilon_0^2 \Delta_r } ,
\end{equation}
we get
\begin{equation}
    ( \tfrac{d r}{d\phi} )^2
    = \tfrac{ \varepsilon_0^4 \Delta_r^2 }{ f^2 \Delta^\ell } (
        \tfrac{1}{b^2}
        - \tfrac{f^2}{ \varepsilon_0^2 \Delta_r }
        + 2 \tfrac{f}{J^2} \mathcal{L}
    ) = V_{eff} .
\end{equation}
The effective potential $V_{eff}$ is reduced to
\begin{equation}
    V_{eff} = \tfrac{ \varepsilon_0^4 \Delta_r^2 }{ f^2 \Delta^\ell } (
        \tfrac{1}{b^2}
        - \tfrac{f^2}{ \varepsilon_0^2 \Delta_r }
    )
\end{equation}
for photons $\mathcal{L} = 0$, where $b = \tfrac{J}{E}$ is the impact parameter.
Solving $V_{eff} = V_{eff , r} = 0$, we find the circular orbit of photons, $r = \pm q r_0$.
The positive value corresponds to Universe 1, and the negative value to Universe 2.
The impact parameter of the photon in circular orbit is
\begin{equation}
\label{impact parameter}
    b_c = \varepsilon_0 \left. \tfrac{\sqrt{\Delta_r}}{f} \right\vert_{ r = \pm q r_0 }
\end{equation}
for both universes.

\begin{figure}
    \centering
    \includegraphics[width=1\linewidth]{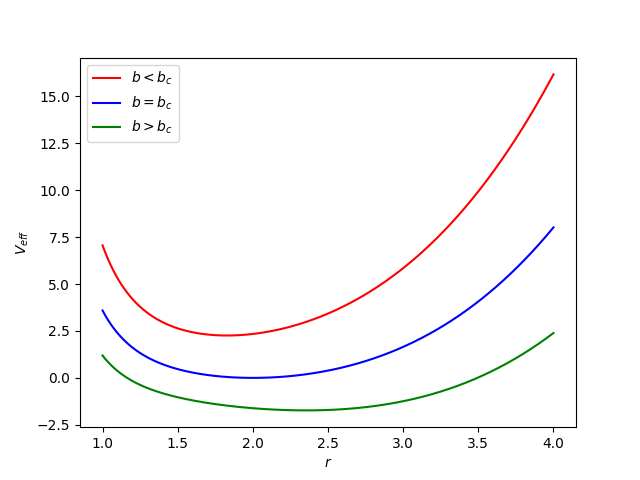}
    \caption{Graphs of $V_{eff} (r)$ for three values of the impact parameter $b$: $b_c - 0.03$, $b_c$ and $b_c + 0.03$.
    To plot them, we set $q = 2$, $r_0 = 1$ and $\theta = \tfrac{\pi}{2}$.}
    \label{fig: eff pot}
\end{figure}

\begin{figure}
    \centering
    \includegraphics[width=\linewidth, scale=1]{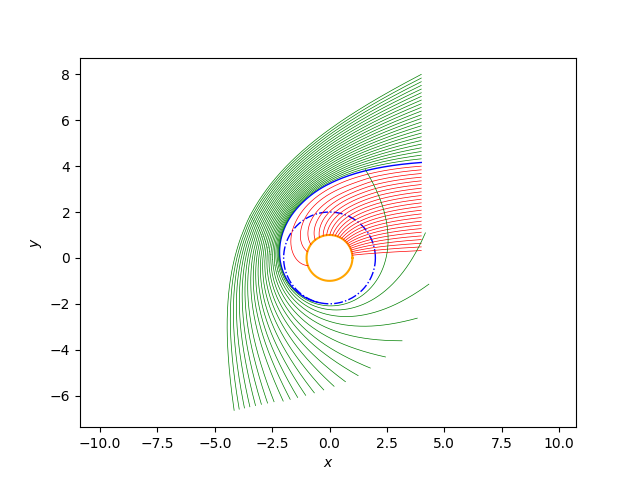}
    \caption{Behavior of geodesics with different values of the impact parameter $b$.
    To plot them, we set $q = 2$, $r_0 = 1$ and $\theta = \tfrac{\pi}{2}$.
    Geodesics with $b < b_c$, $b = b_c$ and $b > b_c$ are shown in red, blue, and green, respectively.}
    \label{fig: geodesics}
\end{figure}

In Figure \ref{fig: eff pot}, we show the graphs of $V_{eff}$ for different values of the impact parameter $b$: $b_c - 0.03 = 0.21424621627637944$, $b_c = 0.24424621627637944$ and $b_c + 0.03 = 0.27424621627637946$.
To plot them, we set $q = 2$, $r_0 = 1$ and $\theta = \tfrac{\pi}{2}$.
Observe that the red curve does not cross the $r$-axis, while the blue curve takes the $r$-axis at $r = 2$. 
The green curve crosses the $r$-axis twice at $r = 1.163715451958726$ and $r = 3.493509720447682$.
However, $r = 3.493509720447682$ is the minimum value that $r$ can take.
This means that geodesics with $b > b_c$ do not reach the wormhole throat, as seen in Figure \ref{fig: geodesics}.
Geodesics with $b = b_c$ also do not reach the wormhole throat because they arrive at $r = 2$ and are trapped in a circular orbit.
Only geodesics with $b < b_c$ cross the wormhole throat to the other universe.

\begin{figure}
    \centering
    \includegraphics[width=\linewidth, scale=1]{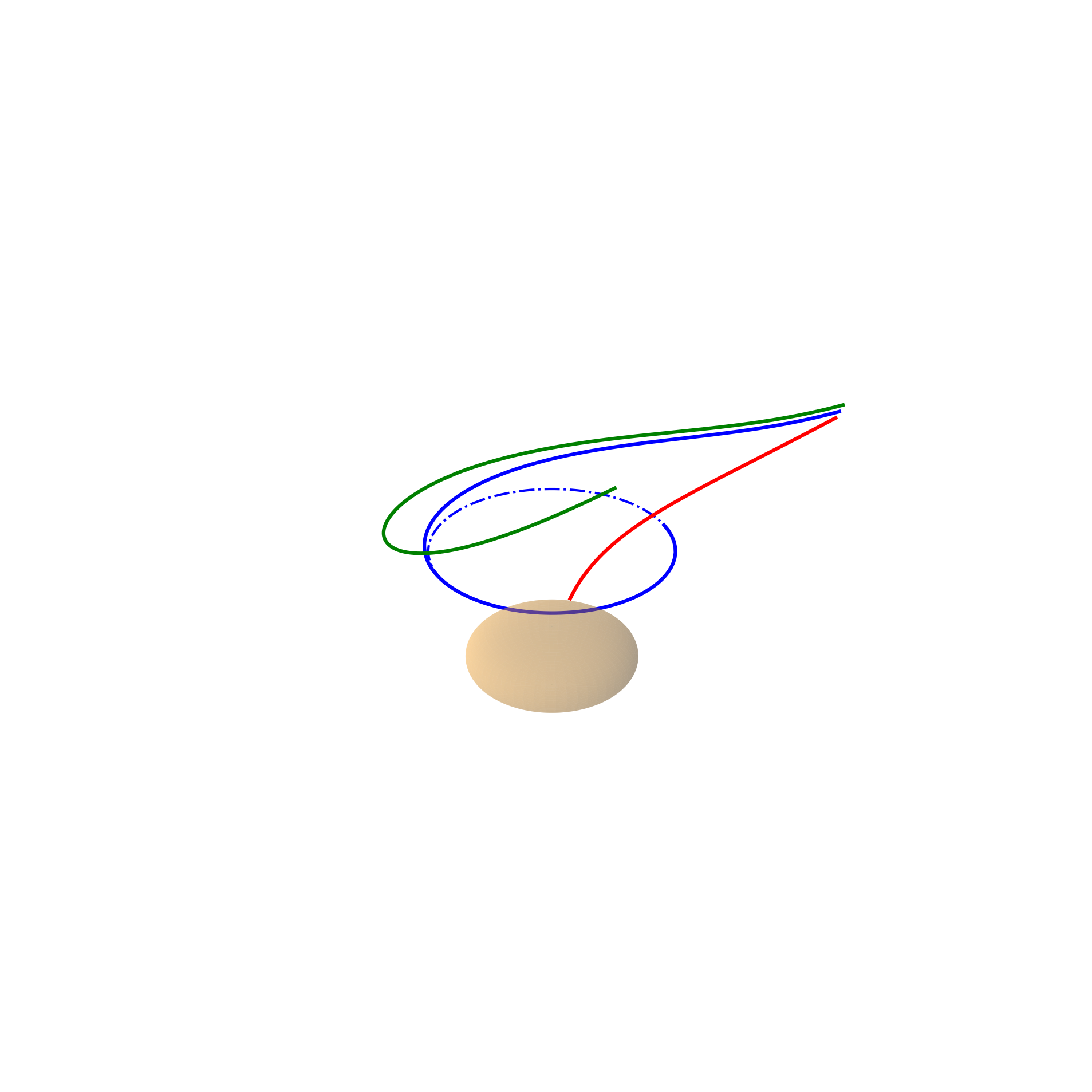}
    \caption{Evolution of geodesics with distinct values of the impact parameter $b$.
    To plot them, we set $q = 2$, $r_0 = 1$ and $\theta = \tfrac{\pi}{4}$.
    Geodesics with $b < b_c$, $b = b_c$ and $b > b_c$ are shown in red, blue, and green, respectively.}
    \label{fig: geodesics 3D}
\end{figure}

Similarly, this occurs for other values of $\theta$, as can be seen in Figure \ref{fig: geodesics 3D}.
Observe that the red geodesic ($b < b_c$) arrives at the wormhole throat, while the blue geodesic ($b = b_c$) is trapped in a circular orbit, and the green geodesic ($b > b_c$) moves away from the wormhole throat.
To plot these geodesics, we set $q = 2$, $r_0 = 1$ and $\theta = \tfrac{\pi}{2}$.

%%%%%%%%%%%%%%%%%%%%%%%%%%%%%%%%
\section{Energy conditions}    %
%%%%%%%%%%%%%%%%%%%%%%%%%%%%%%%%
\label{section: energy conditions}

In general, when we use the cut-and-paste method to construct a wormhole, the derivatives of the metric tensor exhibit a jump singularity at the throat, even when we use vacuum solutions.
This occurs in our case.
The discontinuity induces an effective surface density at the throat, which can be quantified by the surface stress-energy tensor \cite{Israel:1966rt, LorentzianWormholes}, defined as follows:
\begin{equation}
    S_{\mu \nu} =
    \kappa_{\mu \nu} - \kappa h_{\mu \nu} ,
\end{equation}
where
\begin{equation}
    \kappa_{\mu \nu} = K_{\mu \nu}^+ - K_{\mu \nu}^-
\end{equation}
is the discontinuity in the second fundamental form:
\begin{equation}
    K_{\mu \nu}^\pm = \tfrac{1}{2} ( \nabla_\mu^\pm n_\nu + \nabla_\nu^\pm n_\mu ) ,
\end{equation}
$n_\mu$ is the unit vector normal to the shell and
\begin{equation}
    h_{\mu \nu} = g_{\mu \nu} - n_\mu n_\nu
\end{equation}
is the projection tensor.
For our metric, we have
\begin{align}
    S_t^t
&   =\frac{r_G \sin^2 \theta}{4 \pi \Delta_\theta (r_G) n_r } ,
\\  S_\theta^\theta
&   = \frac{r_G}{4 \pi \Delta_r (r_G) n_r} ,
\\  S_\phi^\phi
&   = r_G \frac{\Delta_r (r_G) + r_G^2 \sin^2 \theta}{4 \pi \Delta_r (r_G) \Delta_\theta (r_G) n_r} ,
\end{align}
where $n_r = n_r^+ (r_G) = n_r^- (r_G)$ and $n_r^\pm = \sqrt{\frac{\Delta^\ell}{f_\pm}}$.

In order to check whether the energy conditions are satisfied, we must divide our analysis into two parts.
The first part considers the region outside the throat.
Since the spacetime is vacuum outside the throat, the energy-momentum tensor is zero.
Therefore, the null, weak, dominant, and strong energy conditions are trivially satisfied in this region.

In the second part, we evaluate the energy conditions at the throat.
The null energy condition (NEC) is equivalent to the following inequalities:
\begin{align}
    -S_t^t + S_\theta^\theta \geq 0
    \text{ and } -S_t^t + S_\phi^\phi \geq 0 .
\end{align}
The weak energy condition (WEC) is satisfied if the following inequalities hold:
\begin{align}
    -S_t^t \geq 0
    \text{ and } -S_t^t + S_\theta^\theta \geq 0
    \text{ and } -S_t^t + S_\phi^\phi \geq 0 .
\end{align}
Finally, the strong energy condition (SEC) states that
{\footnotesize
\begin{align}
    -S_t^t + S_\theta^\theta + S_\phi^\phi \geq 0
    \text{ and } -S_t^t + S_\theta^\theta \geq 0
    \text{ and } -S_t^t + S_\phi^\phi \geq 0 .
\end{align} 
}
For our metric, we have
\begin{align}
\label{nec}
    -S_t^t + S_\theta^\theta
&   = r_G \frac{r_G^2 \cos^2 \theta + r_0^2 \cos 2\theta}{4 \pi \Delta_r (r_G) \Delta_\theta (r_G) n_r}
\\  -S_t^t + S_\phi^\phi
&   = \frac{r_G}{4 \pi \Delta_r (r_G) n_r} > 0
\\  -S_t^t + S_\theta^\theta + S_\phi^\phi
&   = \frac{2 r_G}{4 \pi \Delta_r (r_G) n_r} > 0
\end{align}
We set $\theta = \tfrac{\pi}{2}$ in Eq. \eqref{nec}.
Then $-S_t^t + S_\theta^\theta < 0$, so that the NEC, WEC, and SEC do not hold.

%%%%%%%%%%%%%%%%%%%%%%%%%
\section{Tidal Forces}  %
%%%%%%%%%%%%%%%%%%%%%%%%%
\label{section: tidal forces}

Employing the framework presented in \cite{Cita:CarrollBook} and using the vacuum condition $R_{\mu\nu}=0$, we can conclude that the Riemann tensor coincides with the Weyl tensor, i.e.\ $R_{\mu\nu\alpha\beta} = C_{\mu\nu\alpha\beta}$. 

On the other hand, the geodesic deviation equation takes the form
\[
\frac{D^2}{d\tau^2} \tensor{\xi}{^{\beta}}
= \tensor{R}{^{\mu}}{_{\alpha \hspace{0.2em} \beta \hspace{0.2em} \sigma}} 
\tensor{u}{^{\alpha}} \tensor{\xi}{^{\beta}} \tensor{u}{^{\sigma}},
\]
where $\tau$ denotes an affine parameter along the fiducial geodesic and $D/d\tau = u^{\mu} \nabla_{\mu}$ represents the covariant derivative along that geodesic. This formulation guarantees that the relative acceleration between neighboring freely falling particles is a tensorial quantity and, therefore, independent of the choice of observer. Therefore, we can consider the projections of $R_{\mu\alpha \nu \beta} u^{\alpha} u^{\beta} = C_{\mu\alpha \nu \beta} u^{\alpha} u^{\beta}$ along the $\{$\emph{preferred spatial or radial $s^\mu$, axial $b^\mu$, and polar directions $e^\mu$}$\}$:
\begin{align*}
    u&=\left(1/\sqrt{f},0,0,0\right)^T, \\
    s&=\left(0,\sqrt{f}/\Delta^{\ell /2},0,0\right)^T, \\
    e&=\left(0,0,\sqrt{f/\Delta_r}/\Delta^{\ell /2},0\right)^T, \\
    b&=\left(0,0,0,\sqrt{f/\Delta_r}/\sin{\theta}\right)^T.
\end{align*}
In this solution, the Newman–Penrose null tetrad obtained from a diagonal orthonormal frame coincides with the observer-adapted geometric NP tetrad constructed from $(u^\mu,s^\mu,e^\mu,b^\mu)$. This follows from the spacetime’s static nature and the fact that $g_{tt}$ depends only on the radial coordinate. The unit four-velocity of static observers is $u^\mu\propto\partial_t$, and their proper acceleration $a_\mu$ is purely radial. The remaining spatial vectors $b^\mu$ and $e^\mu$ are determined by axial symmetry and orthogonality, aligning with $\partial_\phi$ and $\partial_\theta$, respectively. Thus the geometrically selected orthonormal frame $(u,s,e,b)$ is parallel to the coordinate-adapted diagonal frame $(\partial_t,\partial_r,\partial_\theta,\partial_\phi)$, and the associated null vectors $l^\mu=(u^\mu+s^\mu)/\sqrt{2}$ and $n^\mu=(u^\mu-s^\mu)/\sqrt{2}$ coincide in both constructions, up to normalization.

And the proyections $E_{MN}=C_{\mu\alpha \nu \beta}M^{\mu} N^{\nu} u^{\alpha} u^{\beta}$ where $M,N \in \{s,e,b\}$ are:
{\small
\begin{subequations}\label{Fuerzas de Marea}
    \begin{align}
        E_{ss}=&\frac{f q r_0}{8\, \Delta _{\theta }^{\frac{q^2}{4}+2}} \Delta _r^{\frac{q^2}{4}-1} \bigg(r_0^2 \cos ^2(\theta ) ((q^2-4) r+4 q r_0)\notag \\
        &-r ((q^2+4) r_0^2-4 q r_0 r+8 r^2) \bigg), \\
        E_{ee}=&-\frac{f q r_0}{8\, \Delta _{\theta }^{\frac{q^2}{4}+2}} \Delta _r^{\frac{q^2}{4}-1} \bigg(q r_0^2 \cos ^2(\theta ) (q r+2 r_0) \notag \\
        &-r ((q^2+4) r_0^2-2 q r_0 r+4 r^2) \bigg), \\
        E_{bb}=&-\frac{f q r_0}{4\, \Delta _{\theta }^{\frac{q^2}{4}+1}}  \left(q r_0-2 r\right)  \Delta _r^{\frac{q^2}{4}-1},\\
        E_{se}=&E_{es}=\frac{f q\, \Delta _r^{\frac{1}{4} \left(q^2-2\right)}}{8\, \Delta_{\theta }^{\frac{q^2}{4}+2}}  \left(q^2+4\right) r_0^3 \sin{\theta}  \cos{\theta} 
    \end{align}
\end{subequations}
}
and the remaining ones disappear.

Let's begin by remembering that 
\begin{itemize}
\item $s$ is the \textbf{radial} unit direction,
\item $e$ is the \textbf{polar} unit direction,
\item $b$ is the \textbf{azimuthal}.
\end{itemize}

\paragraph{Azimuthal and mixed polar-radial tidal forces.} By analysing \eqref{Fuerzas de Marea}, we observe that $E_{bb}=0 \quad\Longleftrightarrow\quad r=r_{G}=\frac{q r_0}{2}$, and that $E_{se}\propto \sin\theta\cos\theta$. Consequently, $E_{se}=0 \quad\text{for}\quad \theta=0,\;\frac{\pi}{2},\;\pi$. In other words, at the wormhole throat the azimuthal tidal force vanishes, and the mixed radial–polar tidal component is zero along both the polar axis and the equatorial plane.
Since $E_{bb}\propto(q r_0-2r)$, the azimuthal tidal component flips sign as one crosses the throat: for $r>r_{G}$ we have $(q r_0-2r)<0$, which implies $E_{bb}>0$, whereas for $r<r_{G}$ we obtain $(q r_0-2r)>0$, leading to $E_{bb}<0$. Consequently, the behavior of a small separation in the $\phi$-direction switches from focusing/compression to stretching/defocusing, or vise versa, depending on the sign convention.

\paragraph{Radial and polar tidal forces.} Because we have $E_{bb}(r_{G},\theta)=0$ for every $\theta$, and the vacuum tidal tensor is traceless, satisfying $E_{ss}+E_{ee}+E_{bb}=0$, it follows that
\[
E_{ee}(r_{G},\theta)=-E_{ss}(r_{G},\theta)\qquad(\forall\,\theta).
\]
Thus, at the throat, any radial tidal effect is precisely balanced by an equal and opposite polar effect: radial stretching is accompanied by polar compression (and vice versa).
On the other hand, we have \(E_{ss}(r_{G},0)=E_{ee}(r_{G},0)=0\), and analogously for \(\theta=\pi\). In other words, the radial and polar tidal forces vanish at the throat exclusively along the polar axis. Consequently, this region constitute the most suitable axis for traversing the wormhole.

\begin{figure*}[h]
        \centering
            \includegraphics[width=\textwidth]{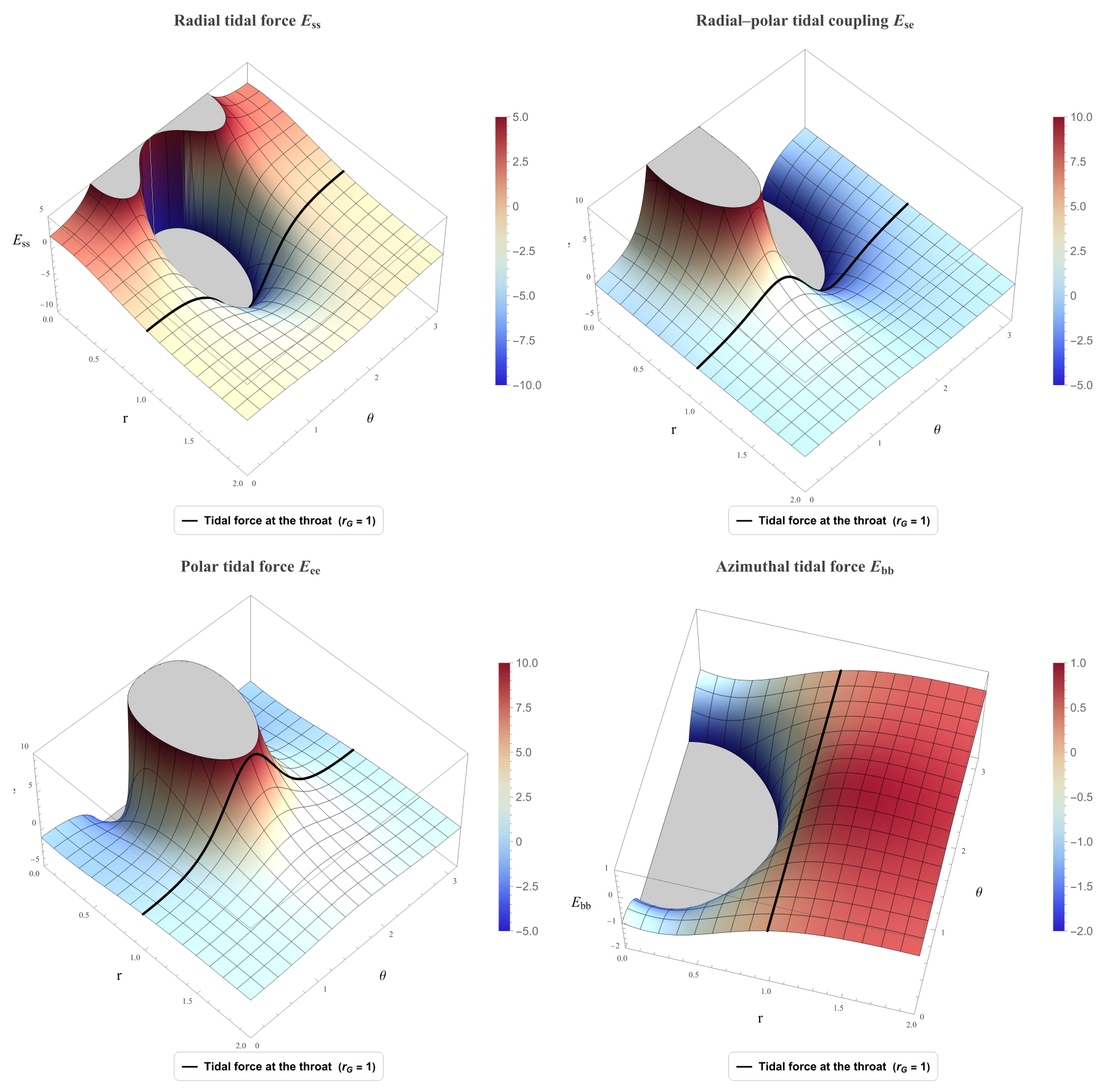}
            \caption{Tidal forces are evaluated by fixing $q=2$ and $r_0=1$ for all the cases of \eqref{Fuerzas de Marea}. The plotted surfaces represent the magnitude ($z$-Axis) of the wormhole, on each surface, a black curve is drawn indicating the tidal forces at the throat $x_G=1$ for any angle. It is evident that within the throat the magnitudes diverge, whereas outside the throat they remain finite and well-behaved.}
            \label{fig:FuerzasMareaq2}
\end{figure*}

An illustration of the tidal forces is shown in Figure \ref{fig:FuerzasMareaq2}, where the case $q=2$ is considered, and the surface depicts the magnitude of each tidal force. The black curve corresponds to the tidal force at the wormhole throat, and we can clearly observe that the throat encompasses all the divergent anomalies of the surfaces. In other words, these anomalies lie inside the throat.

%%%%%%%%%%%%%%%%%%%%%%%%%
\section{Conclusions}   %
%%%%%%%%%%%%%%%%%%%%%%%%%
\label{section: conclusions}

We used the flat subspaces method \cite{Sarmiento-Alvarado2023,Sarmiento-Alvarado2025} to obtain an exact solution $\hat{g}$ to the EFE in vacuum, which depends on three constant parameters, $r_0$, $q$ and $s$.
Since $\hat{g}$ is also valid for negative values of the radial coordinate $r$, we assumed $r \in \mathbb{R}$. In this way, $\hat{g}$ describes a wormhole.

As mentioned earlier, the Komar mass is given by $qr_0/2$, so the parameter $q$, like $r_0$, is directly associated with the wormhole’s mass. In our construction, however, $q \, r_0/2$ also coincides exactly with the throat radius, in other words, the geometric mass and the throat are identical.
We have clarified the role of conical defects. In the extended \(s=0\) geometry, the ring has the same conical excess as in the Gibbons–Volkov solution: \(\delta_{\rm ring}=-2\pi\left(1+\frac{q^2}{4}\right)\). With \(q=2\sigma\), this reproduces \(\delta_{\rm GV}=-2\pi(1+\sigma^2)\). In the cut-and-paste construction, however, this defect lies inside the excised region and does not appear in the final wormhole manifold. In contrast, the \(s\neq0\) sector does not yield a finite axial conical defect but instead develops a stronger axial/polar singularity.
The relation with the scalar and phantom-field model of Gibbons and Volkov is only indirect. Their scalar or phantom configurations enter a solution-generating procedure, but the resulting ring-wormhole geometry is vacuum. Likewise, our \(s=0\) sector is Ricci-flat in the bulk. For \(s\neq0\), \(s\) is not a scalar or phantom charge, but a geometric deformation parameter of the vacuum solution.
Consequently, we set $s = 0$ in our computations and have justified this choice, in particular, the metric $\hat{g}$ is asymptotically flat only in the case $s = 0$. Similarly, this solution is static, with the NUT parameter and the Komar angular momentum both equal to zero.

Furthermore, in th case $s=0$, it has a singularity at $( r, \theta ) = ( 0, \tfrac{\pi}{2} )$, even if $s \neq 0$.
This singularity can be geometrically interpreted as a ring of radius $r_0$.

We assumed that surfaces defined by constant $t$ and $r$ are surfaces of revolution.
However, this occurs only for $\tfrac{r}{r_0} \geq \vert \tfrac{q}{2} \vert$ or $\tfrac{r}{r_0} \leq -\vert \tfrac{q}{2} \vert$.
Hence, we used the cut-paste method to construct wormholes.
In Section \ref{section: throat}, we determine the minimal surfaces.
Only for $\ell = 1$ and $\ell = 2$ were we able to obtain a mathematical expression.

Since \(\ell=1+\frac{q^2}{4}\), the case \(\ell=1\) corresponds to the degenerate limit \(q=0\), namely to the vanishing effective mass limit. In this case, the embedding profile satisfies \(z(\theta)=0\), and the minimal surface reduces to a disk. For \(\ell>1\), the profile is lifted away from the plane \(z=0\), and the throat opens into an oblate-like surface of revolution whose height is controlled by \(\ell\) or by the effective mass parameter \(q\). Thus, increasing \(q\) deforms the disk-like limiting geometry into a non-trivial closed surface of revolution.

In Section \ref{section: embedding surfaces}, we found that surfaces defined by constant $t$ and $\theta \in ( 0, \pi )$ are embedded in the Euclidean space $\mathbb{R}^3$.
The profile curve (Figure \ref{fig: profile curve}) is symmetric with respect to the $\rho$-axis, which implies that the surface (Figure \ref{fig: embbeding surface}) is symmetric with respect to the plane $z = 0$.

The existence of solutions to the equations of motion is established in Section \ref{section: geodesics}.
We focused on solutions with constant $\theta \in (0, \pi)$, such solutions depends on the impact parameter $b = \tfrac{J}{E}$, where $E$ is interpreted as the energy of the system and $J$ as its angular momentum.
The energy $E$ is strictly positive, while $J$ can take any real value.
Only geodesics with an impact parameter $b < b_c$ cross the wormhole throat to the other universe.
Geodesics with $b = b_c$ are trapped in a circular orbit, and geodesics with $b > b_c$ move away from the wormhole throat.

In section \ref{section: energy conditions}, we checked the energy conditions for the wormhole solutions.
As mentioned previously, the cut-and-paste method generates an effective surface stress-energy tensor $S_{\mu \nu}$ due to discontinuities in the derivatives of the metric tensor.
The energy conditions at the throat are expressed in terms of the components of $S_{\mu \nu}$.
We found that outside the throat, the null, weak, and strong energy conditions are satisfied.
However, they are violated at the throat.

Within the Newman–Penrose formalism, employing a Weyl-aligned null tetrad, and based on the asymptotic behavior, we discover that $q$ is related to the mass. For $q \neq 0$, the corresponding spacetime is characterized by Petrov type I. In the asymptotic regime $r \rightarrow \pm \infty$, however, the algebraic structure of the Weyl tensor changes and the geometry tends to a Petrov type D spacetime. 

In the same framework, we further show that the solution is static and does not support gravitational radiation. Due to its invariance under the reflection $r \rightarrow -r$, the geometry admits a consistent interpretation as a wormhole configuration. In the particular case $q = 0$, the metric becomes that of flat Minkowski spacetime, with both the mass and the throat (in the cut-and-paste construction) vanishing.

For this wormhole solution, the analysis of the tidal forces indicates that the safest region for traversability is located in the vicinity of the polar axis. This result is consistent with the findings reported in \cite{Bixano:2025bio}. More importantly, we again conclude that this wormhole configuration satisfies the Wormhole Cosmic Censorship Conjecture (as discussed in \cite{axioms14110831}), since the throat effectively encloses all causal pathologies and trapped surfaces. Indeed, these pathologies and trapped regions are confined to $r < q r_0 / 2$, while the throat is precisely located at $r_G = q r_0 / 2$.

%%%%%%%%%%%%%%%%%%%%%%%%%%%%%
\section*{Acknowledgements} %
%%%%%%%%%%%%%%%%%%%%%%%%%%%%%
This work was partially supported by CONAHCyT M\'exico under grants \mbox{A1-S-8742}, 304001, 376127, 240512,  \mbox{FORDECYT-PRONACES} grant No. 490769 and \mbox{I0101/131/07 C-234/07} of the Instituto Avanzado de Cosmolog\'ia (IAC) collaboration (http://www.iac.edu.mx/).

%%%%%%%%%%%%%%%%%%%%%%%%%%%%%%%%%%%%%%%%%
%\section*{Data Availability Statement}  %
%%%%%%%%%%%%%%%%%%%%%%%%%%%%%%%%%%%%%%%%%

%No new data were created or analysed in this study.

%%%%%%%%%%%%%%%%%%%%%
\begin{appendices}  %
%%%%%%%%%%%%%%%%%%%%%

\section{Differential forms and the Hodge 2-form on the surface S} \label{ApnediceCargasInv}

Let the 2-surface:
\begin{equation}\label{eq:Sx_def}
S_x:\quad t=\text{const},\quad r=\text{const},\quad (\theta,\varphi)\in[0,\pi]\times[0,2\pi).
\end{equation}
For sufficiently large values of $|r|$, this surface takes the form of a topological 2-sphere in the asymptotic region (for both the prolate and oblate coordinate systems). Hence, the limit $r\to\infty$ (or $r\to -\infty$ when a second asymptotic end exists) is equivalent to performing the integration over a large sphere at infinity.
Consider now an antisymmetric 2-form $H$, whose covariant components satisfy $H_{\mu \nu} = -H_{\nu \mu}$. We employ the standard convention
\begin{equation}\label{eq:Hodge_def}
(*H)_{\mu\nu}=\frac12\sqrt{-g}\,\epsilon_{\mu\nu\alpha\beta}\,H^{\alpha\beta},
\qquad
H^{\alpha\beta}=g^{\alpha\gamma}g^{\beta\delta}H_{\gamma\delta},
\end{equation}
where 
\begin{equation*}
    \sqrt{-g}=\frac{\Delta^\ell h\,\Delta_r\,\sin\theta}{f}.
\end{equation*}
Adopting the orientation convention $\epsilon_{tr \theta \varphi}=+1$, we can check (using antisymmetry and suitable index rearrangements) that $\epsilon_{\theta \varphi t r}=+1$. Consequently, from \eqref{eq:Hodge_def} it follows that $(*H)_{\theta \varphi}=\sqrt{-g}\,H^{t r}$. Noting that $g^{rr}=\frac{f}{\Delta^\ell h}$ and $
g^{t t}=-1/f$, we then obtain
\begin{equation}\label{eq:HodgeHMaestra}
(*H)_{\theta \varphi}=\sqrt{-g}\,g^{tt}g^{rr}\,H_{tr}=-\frac{\Delta_r\,\sin\theta}{f} H_{tr}.
\end{equation}
Finally, if $\Omega$ is a 2-form, then on $S_x$ we have $dt = dr = 0$, and therefore
\begin{equation}\label{eq:Integral2FormaSobreSx}
    \int_{S_x}\Omega=\int_0^{2\pi}\!\!d\varphi\int_{0}^{\pi}\!\!d \theta\;(\Omega_{y\varphi})\Big|_{t,r=\text{const}}.
\end{equation}

Thus, to compute the integral of $*H$ over $S_x$, it suffices to consider only the $( *H )_{y\varphi}$ component.
%.......................................

\section{Newman-Penrose Weyl sacalars}\label{Newman-Penrose Weyl Escalares}

{\small
\begin{subequations}
\begin{align}\label{Escalares NPW}
\Psi_4&=\Psi_0= -\frac{f  q \left(q^2+4\right) r r_0^3 \sin ^2(\theta ) \Delta{}^{1-\frac{q^2}{4}}}{16 \Delta _{\theta }^3}, \\
\Psi_3&=-\Psi_1= -\frac{f  q \left(q^2+4\right) r_0^3 \sin (2 \theta ) \Delta ^{-\frac{q^2}{4}-\frac{1}{2}}}{32 \Delta _{\theta }^{3/2} \Delta _r}\\
\Psi_2&= -\frac{f  q r_0 \Delta{}^{1-\frac{q^2}{4}}}{32 \Delta _{\theta }^3} \bigg(r_0^2 r \left(\left(q^2-4\right) \cos (2 \theta )-q^2-12\right) \notag \\ 
&+8 q r_0 r^2+8 q r_0^3 \cos ^2(\theta )-16 r^3\bigg).
\end{align}
\end{subequations}
}

{\small
\begin{subequations}
    \begin{align}\label{Invariantes I J}
    \mathcal{I}&=\frac{q^2 r_0^2 \Delta _{\theta }^{-\frac{q^2}{2}-4} \Delta _r^{\frac{q^2}{2}-2} f^{2} }{1024}
    \bigg(4 \left(q^2+4\right)^2 r^2 r_0^4 \sin ^4(\theta ) \notag \\ 
    &+3 \Big(r_0^2 r \left(\left(q^2-4\right) \cos (2 \theta )-q^2-12\right)+8 q r_0 r^2 \notag \\
    &+8 q r_0^3 \cos ^2(\theta )-16 r^3\Big){}^2+4 \left(q^2+4\right)^2 r_0^4 \sin ^2(2 \theta ) \Delta _r\bigg) , \\
    \mathcal{J}&=-\frac{q^3 r_0^3 \left(q r_0-2 r\right) \Delta _{\theta }^{-\frac{3 q^2}{4}-4} \Delta _r^{\frac{3 q^2}{4}-3} f^{3}}{1024} \cdots \notag \\
    &\cdots\bigg(4 r_0^2 r^2 \left(\left(3 q^2+4\right) \cos (2 \theta )-7 q^2-20\right) \notag \\
    &-2 q r_0^3 r \left(\left(3 q^2-4\right
    ) \cos (2 \theta )-3 q^2-28\right) \notag \\
    &+r_0^4 \Big(\left(q^4+16\right) \cos (2 \theta )-q^4-16 q^2-16\Big)\\
    &+64 q r_0 r^3-64 r^4\bigg).
    \end{align}
\end{subequations}
}

\subsection{Asymptotic behaviour}
Within the Newman–Penrose framework, the leading asymptotic behavior of the Weyl scalars captures the dominant gravitational “Coulomb” contribution, any radiative terms, and higher multipole corrections. For our four-dimensional, vacuum, stationary spacetime, we consider the asymptotic limit as $r \to \pm \infty$, from which we obtain the approximations given in \eqref{Escalares NPW asimptotico}.
\[
    \Delta_r \approx r^2, \qquad \Delta_\theta \approx r^2, \qquad \Delta \approx 1.
\]
and 
\[
    f_{\pm\infty} \approx e^{\pm q \pi/2} 
\]
therefore, \eqref{Escalares NPW} takes the form:
{\small
\begin{subequations}
\begin{align}\label{Escalares NPW asimptotico}
\Psi_4&=\Psi_0= \mp f_{\pm\infty} \frac{ q \ell  r_0^3 \sin ^2\theta }{4 r^5}, \\
\Psi_3&=-\Psi_1= \mp f_{\pm\infty} \frac{  q \ell r_0^3 \sin 2 \theta }{8 r^{5}}\\
\Psi_2&= \pm f_{\pm\infty} \frac{ q r_0}{2 r^3}.
\end{align}
\end{subequations}
}
where $\Xi_{\pm\infty}>0$ denotes a dimensionless constant, $q\in\mathbb{R}$ is a dimensionless parameter, and $r_0$ is a constant.
For the invariants denoted by $\mathcal{I}$ and $\mathcal{J}$, we obtain the following result
{\small
\begin{subequations}\label{Invariantes I J asimptotico}
    \begin{align}
        \mathcal{I} &\approx \frac{3 (f_{\pm\infty})^2 q^2 r_0^2}{4 r^6}, \\
        \mathcal{J} &\approx \mp \frac{ (f_{\pm\infty})^{3} q^3 r_0^3}{8 r^9}.
    \end{align}
\end{subequations}
}
It is straightforward to verify that, upon applying \(27\mathcal{J}^2\) and \(\mathcal{I}^3\), we obtain the expression
\begin{equation}\label{Tipo D asimptotico}
    \mathcal{I}^3=27\mathcal{J}^2 \approx \frac{27 (f_{\pm\infty})^6 q^6 r_0^6}{64 r^{18}}.
\end{equation}

\end{appendices}
%%%%%%%%%%%%%%%%%%%%%%%%%%%%%%%%%%%%%%%%%%
%------------------------------------------------------------

% (opcional) si quieres que las subsecciones salgan como A.a, A.b, ...

\setcounter{secnumdepth}{2}
\renewcommand\thesubsection{\thesection.\alph{subsection}}

\bibliographystyle{elsarticle-harv} 
\bibliography{Bibliografia}

\end{document}